\documentclass[aps, prd, floatfix, nofootinbib, superscriptaddress, twocolumn]{revtex4-2}

\usepackage{amsmath}
\usepackage{amssymb}
\usepackage{amsfonts}
\usepackage{color}
\usepackage{CJKutf8}
\usepackage{dcolumn}
\usepackage{enumerate}
\usepackage{latexsym}
\usepackage{textcomp}

\usepackage[mathscr,scaled=1.15]{urwchancal}
\DeclareFontFamily{OT1}{pzc}{}
\DeclareFontShape{OT1}{pzc}{m}{it}%
{<-> s * [1.15] pzcmi7t}{}
\DeclareMathAlphabet{\mathpzc}{OT1}{pzc}{m}{it}

\usepackage{supertabular}
\usepackage{placeins}
\usepackage{epsfig}
\usepackage{graphicx}
\usepackage{multirow}

\definecolor{purple}{rgb}{0.5,0,0.5}
\definecolor{blue}{rgb}{0.0,0,0.9}
\definecolor{prdblue}{rgb}{0.133,0.118,0.498}
\usepackage[colorlinks=true, pdfstartview=FitV, linkcolor=prdblue, citecolor= prdblue, urlcolor=prdblue]{hyperref}

\newcounter{Afigure}
\newcounter{Atable}

\begin{document}

\begin{CJK}{UTF8}{song}

\title{$\,$\\[-6.5ex]\hspace*{\fill}{\normalsize{\sf\emph{Preprint nos}.\ NJU-INP 064/22, USTC-ICTS/PCFT-22-24}}\\[1.25ex]
Wave functions of $\mathbf{ (I,J^P) = (\tfrac{1}{2},\tfrac{3}{2}^\mp) }$ baryons}

\date{2022 August 25}

\author{Langtian Liu
       $\,^{\href{https://orcid.org/0000-0002-4283-0315}{\textcolor[rgb]{0.00,1.00,0.00}{\sf ID}}}$}
\affiliation{School of Physics, Nanjing University, Nanjing, Jiangsu 210093, China}
\affiliation{Institute for Nonperturbative Physics, Nanjing University, Nanjing, Jiangsu 210093, China}
\author{Chen Chen
       $\,^{\href{https://orcid.org/0000-0003-3619-0670}{\textcolor[rgb]{0.00,1.00,0.00}{\sf ID}}}$}
\email[]{chenchen1031@ustc.edu.cn}
\affiliation{Interdisciplinary Center for Theoretical Study, University of Science and Technology of China, Hefei, Anhui 230026, China}
\affiliation{Peng Huanwu Center for Fundamental Theory, Hefei, Anhui 230026, China}
\author{Craig D.~Roberts%
       $\,^{\href{https://orcid.org/0000-0002-2937-1361}{\textcolor[rgb]{0.00,1.00,0.00}{\sf ID}}}$}
\email[]{cdroberts@nju.edu.cn}
\affiliation{School of Physics, Nanjing University, Nanjing, Jiangsu 210093, China}
\affiliation{Institute for Nonperturbative Physics, Nanjing University, Nanjing, Jiangsu 210093, China}

\begin{abstract}
Using a Poincar\'e-covariant quark+diquark Faddeev equation, we provide structural information on the four lightest $(I,J^P)=(\tfrac{1}{2},\tfrac{3}{2}^\mp)$ baryon multiplets.  These systems may contain five distinct types of diquarks; but in order to obtain reliable results, it is sufficient to retain only isoscalar-scalar and isovector-axialvector correlations, with the latter being especially important.  Viewed with low resolution, the Faddeev equation description of these states bears some resemblance to the associated quark model pictures; namely, they form a set of states related via orbital angular momentum excitation: the negative parity states are primarily $\mathsf P$-wave in character, whereas the positive parity states are $\mathsf D$ wave.  However, a closer look reveals far greater structural complexity than is typical of quark model descriptions, with $\mathsf P$, $\mathsf D$, $\mathsf S$, $\mathsf F$ waves and interferences between them all playing a large role in forming observables.  Large momentum transfer resonance electroexcitation measurements can be used to test these predictions and may thereby provide insights into the nature of emergent hadron mass.
\end{abstract}

\maketitle

\end{CJK}


%
\section{Introduction}
\label{SecIntro}
In working to understand the emergence of baryon mass and structure from quantum chromodynamics (QCD), it is crucial to employ a framework that ensures Poincar\'e invariance of observables \cite{Brodsky:2022fqy} and natural to study colour-singlet three-quark six-point Schwinger functions \cite{Edwards:2011jj, Eichmann:2016yit, Qin:2020rad}.  Baryons appear as poles in such Schwinger functions: the pole location reveals the mass (and width) of a given baryon; and the pole residue is that baryon's Poincar\'e-covariant bound-state wave function.  The nucleon and its excited states appear in the isospin $I=\tfrac{1}{2}$ channel.  Focusing on the proton because it is Nature's only stable hadron, then the lowest-mass spectral feature that can be associated with the $(I,I_z)=(\tfrac{1}{2},+\tfrac{1}{2})$ six-point Schwinger function is an isolated pole on the real axis.  Owing to confinement \cite{Krein:1990sf, Roberts:2007ji, Horn:2016rip}, there is no three-quark continuum; but the related spectral functions exhibit additional structures associated with proton resonances that are attached to poles in the complex plane \cite{Aznauryan:2012ba, Briscoe:2015qia, Brodsky:2020vco, Barabanov:2020jvn}.

Working with the appropriate Schwinger function and using standard techniques, one can derive a Poincar\'e-covariant Faddeev-like equation whose solution provides the masses and wave functions of all baryons with the Poincar\'e-invariant quantum numbers that characterise the channel under consideration.  For instance, the proton and all its radial excitations appear as positive parity solutions of a Faddeev equation derived from the
 $J=\tfrac{1}{2}$ Schwinger function.  The parity partners of these states arise as the negative parity solutions.  Today it is possible to develop a tractable formulation of such problems using the leading-order (rainbow-ladder, RL) truncation in a systematic scheme developed for the continuum bound-state problem \cite{Munczek:1994zz, Bender:1996bb, Qin:2014vya, Binosi:2016rxz}.  The resulting equations have been solved for many baryons \cite{Eichmann:2009qa, Sanchis-Alepuz:2011egq, Sanchis-Alepuz:2014sca, Eichmann:2016hgl, Qin:2018dqp, Qin:2019hgk}.  Although RL truncation does not produce widths for the states, sensible interpretations of the results are available \cite{Eichmann:2008ae, Eichmann:2008ef, Roberts:2011cf}, \emph{viz}.\ the solutions are understood to represent the dressed-quark core of the considered baryon, which is subsequently dressed via meson-baryon final-state interactions \cite{JuliaDiaz:2007kz, Suzuki:2009nj, Ronchen:2012eg, Kamano:2013iva, Garcia-Tecocoatzi:2016rcj}.

\begin{figure}[t]
\centerline{%
\includegraphics[clip, width=0.48\textwidth]{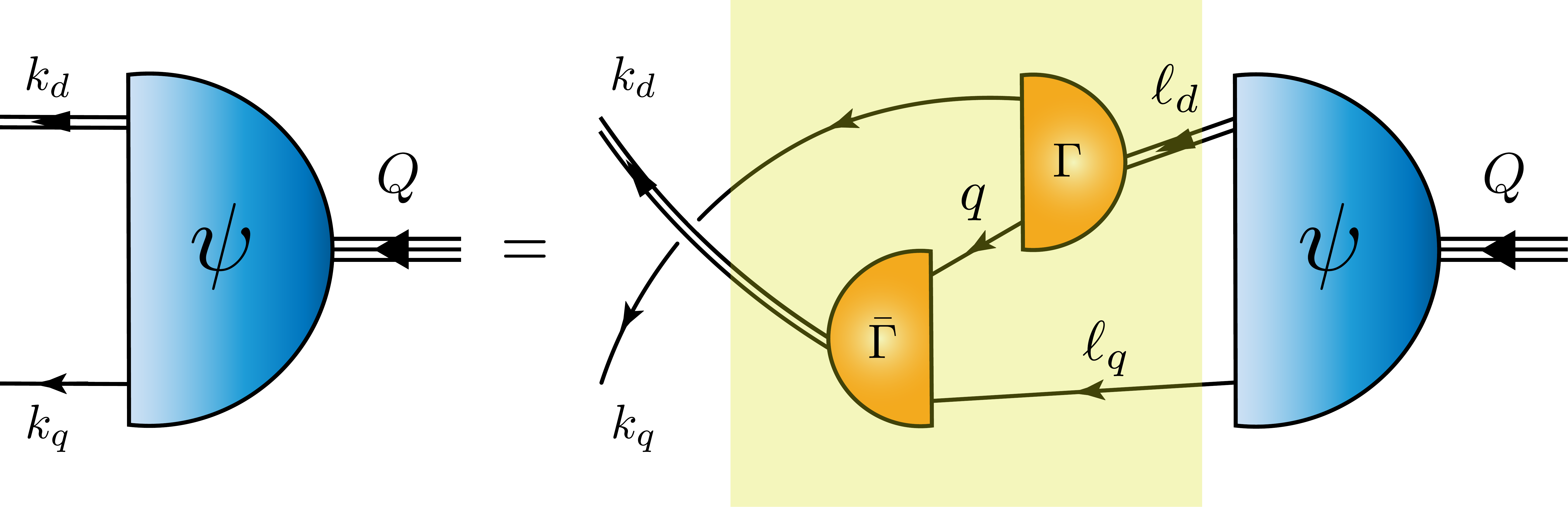}}
\caption{\label{FigFaddeev}
Quark+diquark Faddeev equation, a linear integral equation for the Poincar\'e-covariant matrix-valued function $\psi$, the Faddeev amplitude for a baryon with total momentum $Q=\ell_q+\ell_d=k_q+k_d$. $\psi$ describes the relative momentum correlation between the dressed-quarks and -diquarks. Legend. \emph{Shaded rectangle} -- Faddeev kernel; \emph{single line} -- dressed-quark propagator, $S(q)$; $\Gamma^{J^P}(k;K)$ -- diquark correlation amplitude; and \emph{double line} -- diquark propagator, $D^{J^P}(K)$.
}
\end{figure}

An alternative approach to the Poincar\'e-covariant Faddeev equation exploits the fact that any interaction which provides a good description of ground-state colour-singlet mesons also generates strong colour-antitriplet correlations between any two dressed quarks contained within a hadron \cite{Barabanov:2020jvn}.  This understanding leads to the quark--plus--dynamical-diquark picture of baryon structure, formulated elsewhere \cite{Cahill:1988dx, Burden:1988dt, Reinhardt:1989rw, Efimov:1990uz} and illustrated in Fig.\,\ref{FigFaddeev}.  Here, the kernel is built using dressed-quark and nonpointlike diquark degrees-of-freedom, with binding energy stored within the diquark correlation and additionally generated by the exchange of a dressed-quark, which emerges as one diquark breaks-up and is absorbed into formation of another.  In general, many diquark correlations are possible: isoscalar-scalar, $(I,J^P=0,0^+)$; isovector-axialvector; isoscalar-pseudoscalar; isoscalar-vector; and isovector-vector.  Within a given system, channel dynamics determines the relative strengths of these correlations.

The Faddeev equation in Fig.\,\ref{FigFaddeev} has been used to study the structure of the proton and its lightest $J^P = \tfrac{1}{2}^\pm$ excitations \cite{Chen:2017pse}.  The results indicate that scalar and axialvector diquarks are dominant in the proton and Roper resonance; the associated rest-frame wave functions are primarily $\mathsf S$-wave in character; and the Roper resonance is, at heart, the proton's lightest radial excitation \cite{Burkert:2019bhp, Roberts:2018hpf, Sun:2019aem}.  The predicted presence of axialvector diquarks within the nucleon has far-reaching implications for, \emph{inter alia}, form factors and structure functions \cite{Roberts:2013mja, Cui:2020rmu, Chen:2021guo, Cui:2021gzg, Chang:2022jri, Cheng:2022jxe}.

Regarding $J^P = \tfrac{1}{2}^-$ states, accurate estimates of their masses are obtained by keeping only axialvector diquarks; odd-parity diquarks appear with material strength in the bound-state amplitudes, affecting electroproduction form factors \cite{Raya:2021pyr}; the rest-frame wave functions are dominated by $\mathsf P$-waves, but contain noticeable $\mathsf S$-wave components; and the first excited state, $N(1650)\tfrac{1}{2}^-$, has little of the appearance of a radial excitation of the $N(1535)\tfrac{1}{2}^-$.

So long as rest-frame orbital angular momentum is identified with that existing between dressed-quarks and -diquarks, there are some similarities here with quark model descriptions of $J^P = \tfrac{1}{2}^\pm$ systems.  Notwithstanding that, it should be stressed that in contrast to quark model expectations \cite{Crede:2013kia}, the negative parity states are not simply angular-momentum excitations of the $J^P=\tfrac{1}{2}^+$ ground-states.
It is worth highlighting here that, \emph{inter alia}, any separation of a system's total angular momentum into a sum of constituent orbital angular momentum and spin, $J=L+S$, is frame dependent; hence, in quantum field theory, there is no direct connection between parity -- a Poincar\'e invariant quantity -- and orbital angular momentum.

The Fig.\,\ref{FigFaddeev} Faddeev equation approach was adapted to the analogous low-lying $\Delta$-baryons in Ref.\,\cite{Liu:2022ndb}, revealing that although these states may contain isovector-axialvector and isovector-vector diquarks, the latter contribute little.  The $J^P=\tfrac{3}{2}^+$ systems are the simpler, with some features being similar to quark model pictures, \emph{e.g}., their dominant rest-frame orbital angular momentum component is $\mathsf S$-wave and the $\Delta(1600)\tfrac{3}{2}^+$ looks much like a radial excitation of the $\Delta(1232)\tfrac{3}{2}^+$.
The $J^P=\tfrac{3}{2}^-$ states are more complicated.  In fact, the $\Delta(1940)\tfrac{3}{2}^-$ expresses little of the character of a radial excitation of the $\Delta(1700)\tfrac{3}{2}^-$; and although the rest-frame wave function of the latter is largely $\mathsf P$-wave, matching quark model expectations, this is not true of the $\Delta(1940)\tfrac{3}{2}^-$, whose rest-frame wave function is mostly $\mathsf S$-wave.

An entirely new landscape opens to view when one considers $(I,J^P)=(\tfrac{1}{2},\tfrac{3}{2}^\mp)$ baryons.
%
%
Drawn using the quark model, employing standard notation for SU$(6)$ assignments, these states are interpreted as follows \cite[Sec.\,15]{Workman:2022ynf}:  
\begin{enumerate}[(i)]
\item $N(1520) \tfrac{3}{2}^-$ \ldots\ $(70,1_1^-)$, with constituent-quark orbital angular momentum being $L=1$ and the sum of the three constituent-quark spins being $S=\tfrac{1}{2}$; \label{N1520}
\item $N(1700) \tfrac{3}{2}^-$ \ldots\ $(70,1_1^-)$, $L=1$, $S=\tfrac{3}{2}$;
\item $N(1720)\tfrac{3}{2}^+$ \ldots\  $(56,2_2^+)$, $L=2$, $S=\tfrac{1}{2}$;
\item $N(1900)\tfrac{3}{2}^+$ \ldots\ $(70,2_2^+)$, $L=2$, $S=\tfrac{3}{2}$. \label{N1900}
\end{enumerate}
One sees the parity-partner relationships -- (i), (iii) and (ii), (iv);
but no state is related to another as radial excitation.

In quantum field theory, parity partners are special because all differences between them can be attributed to chiral symmetry breaking; and in the light-quark sector, such symmetry breaking is almost entirely dynamical.  Dynamical chiral symmetry breaking is a corollary of emergent hadron mass (EHM) \cite{Roberts:2020hiw, Roberts:2021xnz, Roberts:2021nhw, Binosi:2022djx, Papavassiliou:2022wrb}; so, linked with confinement in ways that are not yet fully elucidated.  Consequently, experiments that can test predictions made for structural differences between parity partners in the hadron spectrum are valuable \cite{Mokeev:2022xfo, Denisov:2018unj, Aoki:2021cqa}.  Herein, therefore, we present the first predictions for the structure of low-lying $(\tfrac{1}{2},\tfrac{3}{2}^\mp)$ baryons based on the Poincar\'e-covariant Faddeev equation in Fig.\,\ref{FigFaddeev}, the interpretation and validation of which have the potential to shed new light on expressions of EHM in hadron observables.

Our treatment of $(\tfrac{1}{2},\tfrac{3}{2}^\mp)$-baryon bound-state problems is outlined in Sec.\,\ref{secFaddeev}.   This sketch is sufficient because details are provided elsewhere, \emph{e.g}., Refs.\,\cite{Chen:2017pse, Liu:2022ndb, Cloet:2008re, Roberts:2011cf}.
Section~\ref{Solutions} reports predictions for the masses and diquark content of $(\tfrac{1}{2},\tfrac{3}{2}^\mp)$ baryons. 
A detailed picture of $(\tfrac{1}{2},\tfrac{3}{2}^\mp)$ wave functions and interrelationships is drawn in Sec.\,\ref{SecAngular}, with particular attention paid to a rest-frame decomposition of their orbital angular momentum.
Section~\ref{epilogue} presents a summary and perspective.

\section{Bound State Equations}
\label{secFaddeev}
In solving for the masses and structure of $(\tfrac{1}{2},\tfrac{3}{2}^\mp)$ baryons, we aim for unification with existing analyses of $(\tfrac{1}{2},\tfrac{1}{2}^\pm)$, $(\tfrac{3}{2},\tfrac{3}{2}^\mp)$ states \cite{Chen:2017pse, Liu:2022ndb}.  Hence, insofar as possible, the formulations therein are preserved, \emph{e.g}.:
we assume isospin symmetry throughout;
the diquark correlation amplitudes, $\Gamma^{J^P}$, are similar;
the light-quark and diquark propagators, $S$, $\Delta^{J^P}$, are unchanged -- see Ref.\,\cite[Appendix]{Liu:2022ndb};
and the effective masses of the relevant diquark correlations are (in GeV)
\begin{equation}
\label{diquarkmasses}
\begin{array}{cccc}
m_{{[ud]}_{0^+}} & m_{\{uu\}_{1^+}} & m_{{[ud]}_{0^-}} & m_{{[ud]}_{1^-}} \\
0.8 & 0.9 & 1.3 & 1.4
\end{array}\,.
\end{equation}
The mass splitting between diquarks of opposite parity is commensurate with that in the $\rho$-$a_1$ complex  \cite{Workman:2022ynf}.
Moreover, following Ref.\,\cite{Liu:2022ndb}, we emulate Ref.\,\cite[Sec.\,4.1.4]{Yin:2021uom} in electing not to include the $g_{\rm DB}$ channel-coupling suppression-factor discussed in Ref.\,\cite[Sec.\,II.E]{Chen:2017pse} because, as will become apparent, negative parity diquarks play only a modest role in $(\tfrac{1}{2},\tfrac{3}{2}^\mp)$ baryons.
Finally, we neglect isovector-vector diquark correlations because their couplings into all systems are negligible, \emph{e.g}., it is just 24\% of the isoscalar-vector strength and 1\% of that associated with the isoscalar-scalar correlation.

For $(\tfrac{1}{2},\tfrac{3}{2}^P)$ baryons represented as quark+diquark bound states, the full Faddeev amplitude has the form
\begin{align}
\underline{\psi} & = \psi_1 + \psi_2 + \psi_3\,,
\end{align}
where the subscript identifies that quark which is not participating in a diquark correlation and $\psi_{1,2}$ are obtained from $\psi_3=:\psi$ by a cyclic permutation of all quark labels \cite{Cahill:1988dx}.  Focusing on the positive electric charge state without loss of generality, one may write
\begin{align}
\nonumber
\psi^P(k_i,&\alpha_i,\sigma_i) \\ 
%
%
\nonumber
& = [\Gamma^{0^+}(l;K)]^{\alpha_1 \alpha_2}_{\sigma_1 \sigma_2} \, \Delta^{0^+}(K) \,[ \, {\mathpzc S}_{\rho}^P(k;Q) u_\rho(Q)]^{\alpha_3}_{\sigma_3} \\
\nonumber
& \quad +   [{\tt t}^j \Gamma^{1^+}_\mu] \, \Delta_{\mu\nu}^{1^+}\,
[{\mathpzc A}_{\nu\rho }^{j P}(k;Q) u_\rho(Q)] \\
\nonumber
& \quad+   [\Gamma^{0^- }]\, \Delta^{0^-}\, [{\mathpzc P}_{\rho}^P(k;Q) u_\rho(Q)] \\
& \quad+   [\Gamma^{1^-}_\mu] \,\Delta_{\mu\nu}^{1^-}\,
[{\mathpzc V}_{\nu\rho}^P(k;Q) u_\rho(Q)] \,,
\label{FaddeevAmp}
\end{align}
where
$(k_i,\sigma_i,\alpha_i)$ are the momentum, spin and isospin labels of the quarks constituting the bound state;
$Q=k_1 + k_2 + k_3=k_d+k_q$ is the total momentum of the baryon, $Q^2 = \hat Q^2 M^2 = - M^2$, $M$ is the baryon's mass;
$l=(k_1-k_2)/2$, $K=k_1+k_2=k_d$, $k = (-K + 2 k_3)/3$;
${\tt t}^j$, $j=\{+,0\}$, are axialvector diquark isospin matrices, with $j$ summed in Eq.\,\eqref{FaddeevAmp};
and $u_\rho(Q)$ is a Rarita-Schwinger spinor, in which we have here suppressed the spin-projection label.

In Eq.\,\eqref{FaddeevAmp},
{\allowdisplaybreaks
\begin{subequations}
\label{ScalarFunctionsFA}
\begin{align}
{\mathpzc S}_{\rho}^P(k;Q) & = \sum_{i=1}^2 {\mathpzc v}^{i}_{0^+}(k;Q) {\mathpzc G}^{P} {\mathpzc X}^i_\rho(k;Q) \,, \\
{\mathpzc A}_{\nu\rho}^{jP}(k;Q) & = \sum_{i=1}^8 {\mathpzc v}_{1^+}^{ji}(k;Q) {\mathpzc G}^{P} {\mathpzc Y}^i_{\nu\rho}(k;Q)\,, \\
{\mathpzc P}_{\rho}^P(k;Q) & = \sum_{i=1}^2 {\mathpzc v}_{0^-}^i(k;Q) {\mathpzc G}^{-P} {\mathpzc X}^i_\rho(k;Q) \,,\\
{\mathpzc V}_{\nu\rho}^{P}(k;Q) & = \sum_{i=1}^8 {\mathpzc v}^{i}_{1^-}(k;Q) {\mathpzc G}^{-P} {\cal Y}^i_{\nu\rho}(k;Q) \,,
\end{align}
\end{subequations}
where ${\mathpzc G}^{+(-)} = {\mathbb I}_{\rm D} (i\gamma_5)$ and, with
$T_{\mu\nu}=\delta_{\mu\nu}+\hat Q_\mu \hat Q_\nu$,
$\gamma_\mu^\perp = T_{\mu\nu} \gamma_\nu$,
$k_\mu^\perp = T_{\mu\nu} k_\nu$,
$\hat k_\mu^\perp \hat k_\mu^\perp  = 1$,
\begin{subequations}
\label{XYbasis}
\begin{align}
{\mathpzc X}^1_\rho(k;Q) & = i\surd 3\,\hat k_\rho^\perp\gamma_5\,,\\
{\mathpzc X}^2_\rho(k;Q) & = i\gamma\cdot \hat k^\perp\,{\mathpzc X}^1_\rho(k;Q) \,,\\
{\mathpzc Y}^1_{\nu\rho}(k;Q) & = \delta_{\nu\rho}{\mathbb I}_{\rm D}\,,\\
{\mathpzc Y}^2_{\nu\rho}(k;Q) & = \tfrac{i}{\surd 5}[2 \gamma_\nu^\perp \hat k^\perp_\rho - 3 \delta_{\nu\rho}
\gamma\cdot \hat k^\perp]\,,\\
{\mathpzc Y}^3_{\nu\rho}(k;Q) & = -i\gamma^\perp_\nu \hat k^\perp_\rho\,, \\
{\mathpzc Y}^4_{\nu\rho}(k;Q) & = \surd 3 \hat Q_\nu \hat k_\rho^\perp \,,\\
{\mathpzc Y}^5_{\nu\rho}(k;Q) & = 3 \hat k^\perp_\nu \hat k^\perp_\rho - \delta_{\nu\rho} - \gamma^\perp_\nu \hat k^\perp_\rho \gamma\cdot\hat k^\perp\,,\\
{\mathpzc Y}^6_{\nu\rho}(k;Q) & = \gamma_\nu^\perp \hat k_\rho^\perp \gamma\cdot\hat k^\perp\,, \\
{\mathpzc Y}^7_{\nu\rho}(k;Q) & = -i \gamma\cdot \hat k^\perp {\mathpzc Y}^4_{\nu\rho}(k;Q) \,, \\
{\mathpzc Y}^8_{\nu\rho}(k;Q) & = \tfrac{i}{\surd 5} [
\delta_{\nu \rho} \gamma\cdot \hat k^\perp \nonumber \\
& \qquad + \gamma^\perp_\nu \hat k^\perp_\rho -5 \hat k^\perp_\nu \hat k^\perp_\rho \gamma\cdot \hat k^\perp]\,.
\end{align}
\end{subequations}
Further, with $\Lambda_+(Q) = (M-i\gamma\cdot Q )/(2 M)$,
\begin{subequations}
\begin{align}
& \frac{1}{2 M} \sum_{r=-3/2}^{3/2} u_\rho(Q;r) \bar u_\mu(Q;r)
= \Lambda_+(Q) R_{\rho\mu}\,,\\
R_{\rho\mu}& =
\delta_{\rho\mu} {\mathbb I}_{\rm D}  \nonumber \\
& \quad -\tfrac{1}{3}\gamma_\rho \gamma_\mu
+\tfrac{2}{3} \hat Q_\rho \hat Q_\mu {\mathbb I}_{\rm D}
- \tfrac{i}{3} [ \hat Q_\rho \gamma_\mu - \hat Q_\mu \gamma_\rho]\,.
\end{align}
\end{subequations}
Details of our Euclidean metric conventions are presented elsewhere \cite[Appendix\,B]{Segovia:2014aza}.
}

Working with the amplitude in Eq.\,\eqref{FaddeevAmp}, straightforward algebra translates the Fig.\,\ref{FigFaddeev} Faddeev equation into a linear, homogeneous matrix equation for the coefficient functions that may figuratively be written as follows:
\begin{align}
& \left[
\begin{array}{l}
{\mathpzc S}_{\rho}^P(k;Q) \\
{\mathpzc A}_{\mu\rho }^{j P}(k;Q)\\
{\mathpzc P}_{\rho}^P(k;Q) \\
{\mathpzc V}_{\mu\rho}^P(k;Q)
\end{array}\right] u_\rho \nonumber \\
& = 2 \int \frac{d^4\ell}{(2\pi)^4}
\left[ {\cal K}_{(\mu\nu)}^{(jm)}(k,\ell;Q)\right]
\left[
\begin{array}{l}
{\mathpzc S}_{\rho}^P(k;Q) \\
{\mathpzc A}_{\nu\rho }^{m P}(k;Q)\\
{\mathpzc P}_{\rho}^P(k;Q) \\
{\mathpzc V}_{\nu\rho}^P(k;Q)
\end{array}\right] u_\rho\,,
\label{AlgebraFE}
\end{align}
in which, \emph{e.g}., the ${\mathpzc A}_{\mu\rho }^j$--${\mathpzc A}_{\nu\rho }^m$ entry in the kernel matrix is
\begin{align}
 {\cal K}_{\mu\nu}^{jm}&(k,\ell;Q) =
{\tt t}^j\, \Gamma_\alpha^{1^+}\!(k_q-\ell_{d}/2;\ell_{d})
S^{\rm T}(\ell_{d}-k_q) \nonumber \\
&\quad \times {\tt t}^m\,\bar\Gamma^{1^+}_\mu\!(\ell_q-k_{d}/2;-k_{d})\,
S(\ell_q)\,\Delta^{1^+}_{\alpha\nu}(\ell_{d})\,,
\end{align}
where the isospin matrices are
\begin{equation}
{\tt t}^+ = \tfrac{1}{\surd 2}(\tau^0 + \tau^3)\,,\;
{\tt t}^0 = \tau^1 \,,
\end{equation}
$\tau^0 = {\rm diag}[1,1]$, $\{\tau^i\,|\, i=1,2,3\}$ are the usual Pauli matrices,
$\ell_q=\ell+Q/3$, $k_q=k+Q/3$, $\ell_{d}=-\ell+ 2Q/3$,
$k_{d}=-k+2Q/3$, and ``T'' denotes matrix transpose.
The kernel matrix draws connections between all diquark correlations in the complete amplitude, Eq.\,\eqref{FaddeevAmp}, \emph{e.g}., ${\mathpzc S}\to {\mathpzc S}, {\mathpzc A}, {\mathpzc P}, {\mathpzc V}$;
and when written explicitly for all scalar functions in Eqs.\,\eqref{ScalarFunctionsFA}, $[{\cal K}_{(\mu\nu)}^{(jm)}]$ is a $28\times 28$ matrix.  This is reduced to $20\times 20$ if one exploits isospin symmetry for the axialvector diquarks.

\begin{figure*}[!t]
\centerline{%
\includegraphics[clip, width=0.98\textwidth]{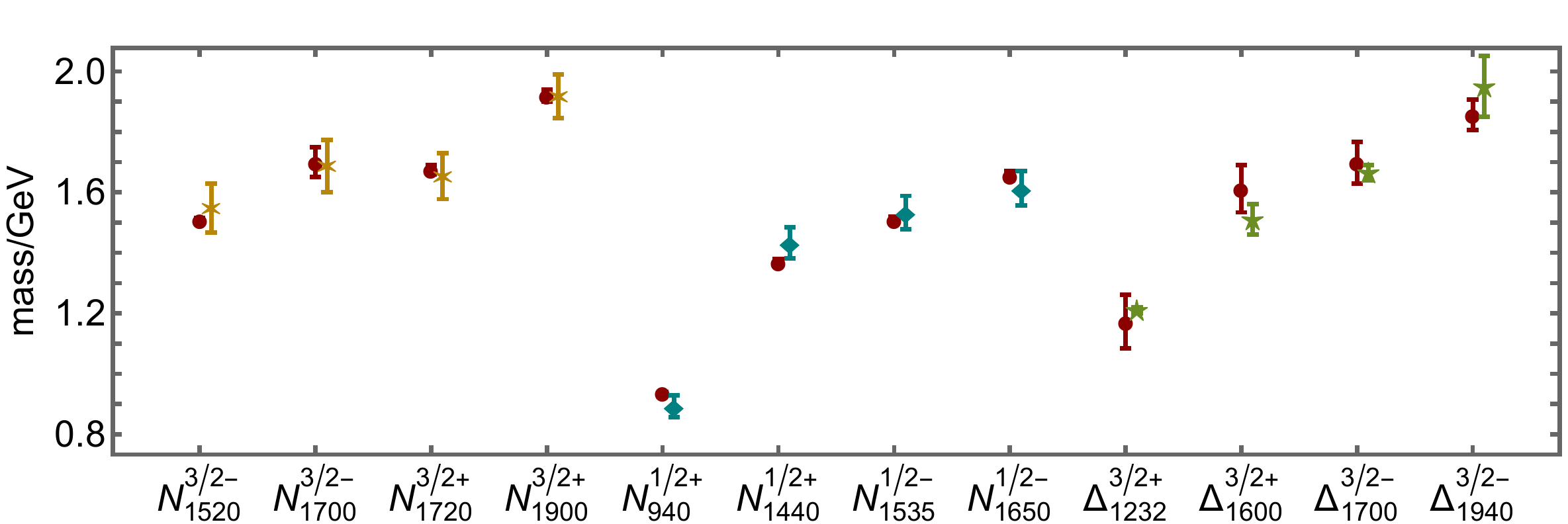}}
\caption{\label{MassCompare}
Real part of empirical pole position for each identified baryon \cite{Workman:2022ynf} (gold asterisk) compared with:
calculated masses in Table~\ref{N32Masses} (red circles) after subtracting $\delta^{N_{3/2}}_{\rm MB}=0.13\,$GeV from each;
calculated masses in Ref.\,\cite[Eq.\,(15)]{Chen:2017pse} (teal diamonds) after subtracting $\delta^{N_{1/2}}_{\rm MB}=0.30\,$GeV;
and
calculated masses in Ref.\,\cite[Table~II]{Liu:2022ndb} (green five-pointed stars) after subtracting $\delta^{\Delta_{3/2}}_{\rm MB}=0.17\,$GeV.
All theory values are drawn with an uncertainty that reflects a $\pm 5$\% change in diquark masses, Eq.\,\eqref{diquarkmasses}.
}
\end{figure*}

The diquark correlation amplitudes are explained in Ref.\,\cite[Eq.\,(1)]{Chen:2017pse}, but it is worth repeating some of the information here:
{\allowdisplaybreaks
\begin{subequations}
\label{qqBSAs}
\begin{align}
\Gamma^{0^+}(k;K) & = g_{0^+} \, i\gamma_5 C\, i\tau^2 \,\vec{H} \,{\mathpzc F}(k^2/\omega_{0^+}^2) \,, \\
%
%
{\tt t}^j \Gamma_\mu^{1^+}(k;K)
    & = i g_{1^+} \, \gamma_\mu C \, {\tt t}^j \, \vec{H} \,
    {\mathpzc F}(k^2/\omega_{1^+}^2)\,,\\
\Gamma^{0^-}(k;K) & =  g_{0^-} \, C\, i\tau^2 \,\vec{H} \,{\mathpzc F}(k^2/\omega_{0^-}^2)\,,\\
%
%
\Gamma_\mu^{1^-}(k;K) & = g_{1^-} \, \gamma_\mu \gamma_5 C\, i\tau^2 \,\vec{H} \,{\mathpzc F}(k^2/\omega_{1^-}^2)\,,
%
%
\end{align}
\end{subequations}
where $C=\gamma_2\gamma_4$ is the charge conjugation matrix;
$\vec{H} = \{i\lambda_c^7, -i\lambda_c^5,i\lambda_c^2\}$, with $\{\lambda_c^k,k=1,\ldots,8\}$ denoting Gell-Mann matrices in colour space, expresses the diquarks' colour antitriplet character; and
${\cal F}(z)=[1-\exp(-z)]/z$.  The correlation widths in Eqs.\,\eqref{qqBSAs} are defined by the related masses \cite[Eq.\,(5)]{Chen:2017pse}: $\omega_{J^P}^2 = m_{J^P}^2/2$.
The amplitudes are canonically normalised \cite[Eq.\,(3)]{Chen:2017pse}, which entails:
\begin{equation}
\begin{array}{l}
g_{0^+} = 14.8\,,\;
g_{1^+} = 12.7\,,\;
g_{0^-} = 6.59\,,\; \\
g_{1^-}^{I=0} = 3.27\,,\;
g_{1^-}^{I=1} = 1.59\,.
\end{array}
\end{equation}
It is the coupling-squared which appears in the Faddeev kernel, so one should expect negative-parity diquarks to play a limited role in the Faddeev amplitudes.

Using the information above, and standard diquark and quark dressed-propagators -- Ref.\,\cite[Eqs.\,(4) and Appendix A]{Chen:2017pse}, the masses and Faddeev amplitudes of the ground- and first-excited state in both the positive- and negative-parity $(I,J)=(\tfrac{1}{2},\tfrac{3}{2})$ channels can be obtained straightforwardly by solving the Faddeev equation -- Fig.\,\ref{FigFaddeev},  Eq.\,\eqref{AlgebraFE} -- using readily available software \cite{Arpack, SPECTRA}.

Given the importance of orbital angular momentum in the discussion of $(\tfrac{1}{2},\tfrac{3}{2}^P)$ baryons and since it is only when working with the wave function that meaningful angular momentum decompositions become available, we record here that the (unamputated) Faddeev wave function is recovered from the amplitude by reattaching the quark and diquark propagator legs independently and appropriately to each term in Eq.\,\eqref{FaddeevAmp}.  Namely, one multiplies Eq.\,\eqref{AlgebraFE} from the left by
\begin{equation}
S(k_q)
\left[\begin{array}{cccc}
\Delta^{0^+}(k_{d}) & 0 & 0 & 0\\
0 & \Delta_{\mu^\prime \mu}^{1^+}(k_{d}) & 0 & 0 \\
0 & 0 & \Delta^{0^-}(k_{d}) & 0\\
0 & 0 & 0 & \Delta_{\mu^\prime \mu}^{1^-}(k_{d})
\end{array}\right]
\end{equation}
to obtain a modified equation for the following wave function:
{\allowdisplaybreaks
\begin{align}
\nonumber
\Psi^P(k_i,\alpha_i, \sigma_i)
%
 & = [\Gamma^{0^+}(l;K)]^{\alpha_1 \alpha_2}_{\sigma_1 \sigma_2} \, [ \underline{\mathpzc S}_{\rho}^P(k;Q) u_\rho(Q)]^{\alpha_3}_{\sigma_3} \\
\nonumber
& \quad +   [\Gamma^{1^+j}_\mu] \,
[\underline{\mathpzc A}_{\mu\rho }^{j P}(k;Q) u_\rho(Q)] \\
\nonumber
& \quad+   [\Gamma^{0^- }]\,  [\underline{\mathpzc P}_{\rho}^P(k;Q) u_\rho(Q)] \\
& \quad+   [\Gamma^{1^-}_\mu] \,
[{\mathpzc V}_{\mu\rho}^P(k;Q) u_\rho(Q)] \,,
\label{FaddeevWF}
\end{align}
where, \emph{e.g}.,
$\underline{\mathpzc A}_{\mu\rho }^{j P}(k;Q) = S(k_q)\Delta_{\mu\nu}^{1^+}(k_d)\,
{\mathpzc A}_{\nu\rho }^{j P}(k;Q)$.
%
%
Each of the functions here has an expansion analogous to that in Eq.\,\eqref{ScalarFunctionsFA}:
\begin{subequations}
\label{ScalarFunctions}
\begin{align}
\underline{\mathpzc S}_{\rho}^P(k;Q) & = \sum_{i=1}^2 {\mathpzc w}^{i}_{0^+}(k;Q) {\mathpzc G}^{P} {\mathpzc X}^i_\rho(k;Q) \,, \\
\underline{\mathpzc A}_{\nu\rho}^{jP}(k;Q) & = \sum_{i=1}^8 {\mathpzc w}_{1^+}^{ji}(k;Q) {\mathpzc G}^{P} {\mathpzc Y}^i_{\nu\rho}(k;Q)\,, \\
\underline{\mathpzc P}_{\rho}^P(k;Q) & = \sum_{i=1}^2 {\mathpzc w}_{0^-}^i(k;Q) {\mathpzc G}^{-P} {\mathpzc X}^i_\rho(k;Q) \,,\\
\underline{\mathpzc V}_{\nu\rho}^{P}(k;Q) & = \sum_{i=1}^8 {\mathpzc w}^{i}_{1^-}(k;Q) {\mathpzc G}^{-P} {\cal Y}^i_{\nu\rho}(k;Q) \,.
\end{align}
\end{subequations}
At this point, one may make the rest-frame angular momentum associations listed in Table~\ref{TabL}.  
}

\begin{table}[t]
\caption{\label{TabL}
Working with the wave function defined in Eq.\,\eqref{FaddeevWF}, decomposed as in Eq.\,\eqref{ScalarFunctions}, and projected into the rest frame, one has the tabulated $J=\tfrac{3}{2}=L+S$ angular momentum decomposition.  The last row lists the associated spectroscopic label, with the $J=\tfrac{3}{2}$ subscript suppressed.
 }
\begin{center}
\begin{tabular*}
{\hsize}
{
l@{\extracolsep{0ptplus1fil}}
c@{\extracolsep{0ptplus1fil}}
c@{\extracolsep{0ptplus1fil}}
c@{\extracolsep{0ptplus1fil}}
c@{\extracolsep{0ptplus1fil}}
c@{\extracolsep{0ptplus1fil}}
c@{\extracolsep{0ptplus1fil}}}\hline
$L$ & $0$ & $1$ & $1$ & $2$ & $2$ &$3$ \\
$S$ & $\tfrac{3}{2}$ & $\tfrac{3}{2}$ & $\tfrac{1}{2}$ & $\tfrac{3}{2}$ & $\tfrac{1}{2}$ &$\tfrac{3}{2}\ $ \\[1ex] \hline
$\Psi^{P}\ $ & ${\mathpzc w}_{1^\pm}^1\ $ & ${\mathpzc w}_{1^\pm}^{2}\ $
& ${\mathpzc w}_{0^\pm}^1$, ${\mathpzc w}_{1^\pm}^{3,4}\ $  & ${\mathpzc w}_{1^\pm}^{5}\ $ & ${\mathpzc w}_{0^\pm}^2$, ${\mathpzc w}_\pm^{6,7}\ $ & ${\mathpzc w}_{1^\pm}^8\ $\\
& $^4{\mathsf S}$ & $^4{\mathsf P}$ & $^2{\mathsf P}$ & $^4{\mathsf D}$ & $^2{\mathsf D}$ & $^4{\mathsf F}$ \\\hline
\end{tabular*}
\end{center}
\end{table}

\section{Faddeev Equation Solutions}
\label{Solutions}
%
%
Solving the Faddeev equation with the full amplitude in Eq.\,\eqref{FaddeevAmp}, one obtains the following masses (in GeV):
\begin{equation}
\label{N32Masses}
\begin{array}{cccc}
N(1520)\tfrac{3}{2}^- & N(1700)\tfrac{3}{2}^- & N(1720)\tfrac{3}{2}^+ & N(1900)\tfrac{3}{2}^+ \\
1.68(8) & 1.82(9) & 1.78(8) & 2.05(7)
\end{array}\,,
\end{equation}
where the indicated uncertainties express the result of a $\pm5$\% change in the diquark masses, Eq.\,\eqref{diquarkmasses}.
As explained elsewhere \cite{Hecht:2002ej, Sanchis-Alepuz:2014wea, Chen:2017pse, Burkert:2019bhp, Liu:2022ndb}, the kernel in Fig.\,\ref{FigFaddeev} omits all contributions that may be linked with meson-baryon final-state interactions, \emph{i.e}., the terms which transform a bare-baryon into the observed state after their inclusion, \emph{e.g}., via dynamical coupled channels calculations \cite{JuliaDiaz:2007kz, Suzuki:2009nj, Ronchen:2012eg, Kamano:2013iva, Garcia-Tecocoatzi:2016rcj}.  The Faddeev amplitudes and masses we obtain should therefore be viewed as describing the \emph{dressed-quark core} of the bound-state, not the completely-dressed, observable object \cite{Eichmann:2008ae, Eichmann:2008ef, Roberts:2011cf}.  That explains why the masses are uniformly too large.

\begin{table}[t]
\caption{\label{qqmassfractions}
{\sf Panel A}. Baryon masses (in GeV) calculated using the indicated diquark correlations: ${\rm all} = 1^+, 0^+,0^-,1^-$.
{\sf Panel B}. Change in a baryon's mass generated by the progressive inclusion of additional diquark correlations, in the order listed.
{\sf Panel C}. Fraction of a given baryon's Faddeev amplitude (FA) contributed by the different diquark correlations, defined in connection with Eq.\,\eqref{FractionFA}.
(In {\sf B} and {\sf C}, the sum of entries in each column is unity.)
 }
\begin{center}
\begin{tabular*}
{\hsize}
{
l@{\extracolsep{0ptplus1fil}}
|c@{\extracolsep{0ptplus1fil}}
|c@{\extracolsep{0ptplus1fil}}
|c@{\extracolsep{0ptplus1fil}}
|c@{\extracolsep{0ptplus1fil}}}\hline
{\sf A}. mass\; & $N(1520)\tfrac{3}{2}^-\ $ & $N(1700)\tfrac{3}{2}^-\ $ & $N(1720)\tfrac{3}{2}^+\ $ & $N(1900)\tfrac{3}{2}^+\ $ \\\hline
$1^+\ $ & $1.84\ $ & $2.06\ $ & $1.90\ $ & $1.98\ $ \\
$0^+\ $ & $1.99\ $ & $2.22\ $ & $1.98\ $ & $2.15\ $ \\
$0^-\ $ & $2.34\ $ & $2.59\ $ & $2.49\ $ & $2.61\ $ \\
$1^-\ $ & $2.49\ $ & $2.72\ $ & $2.44\ $ & $2.62\ $ \\\hline
$1^+, 0^+\ $ & $1.68\ $ & $1.88\ $ & $1.78\ $ & $2.04\ $ \\\hline
$1^+, 0^+,0^-\ $ & $1.68\ $ & $1.88\ $ & $1.78\ $ & $2.05\ $ \\\hline
all\; & $1.68\ $ & $1.82\ $ & $1.78\ $ & $2.05\ $ \\\hline
\end{tabular*}

\medskip

\begin{tabular*}
{\hsize}
{
l@{\extracolsep{0ptplus1fil}}
|c@{\extracolsep{0ptplus1fil}}
|c@{\extracolsep{0ptplus1fil}}
|c@{\extracolsep{0ptplus1fil}}
|c@{\extracolsep{0ptplus1fil}}}\hline
{\sf B}. mass \%\; & $N(1520)\tfrac{3}{2}^-\ $ & $N(1700)\tfrac{3}{2}^-\ $ & $N(1720)\tfrac{3}{2}^+\ $ & $N(1900)\tfrac{3}{2}^+\ $ \\\hline
$1^+\ $ & $91.6\ $ & $89.0\ $ & $94.1\ $ & $96.9\ $ \\
\& $0^+\ $ & $\phantom{9}8.3\ $ & $\phantom{9}8.0\ $ & $\phantom{9}5.8\ $ & $\phantom{9}2.8\ $ \\
\& $0^-\ $ & $\phantom{9}0.1\ $ & $\phantom{9}0.2\ $ & $\phantom{9}0.1\ $ & $\phantom{9}0.3\ $ \\
\& $1^-\ $ & $\phantom{9}0.0\ $ & $\phantom{9}2.8\ $ & $\phantom{9}0.0\ $ & $\phantom{9}0.0\ $ \\\hline
\end{tabular*}

\medskip

\begin{tabular*}
{\hsize}
{
l@{\extracolsep{0ptplus1fil}}
|c@{\extracolsep{0ptplus1fil}}
|c@{\extracolsep{0ptplus1fil}}
|c@{\extracolsep{0ptplus1fil}}
|c@{\extracolsep{0ptplus1fil}}}\hline
{\sf C}. FA \%\; & $N(1520)\tfrac{3}{2}^-\ $ & $N(1700)\tfrac{3}{2}^-\ $ & $N(1720)\tfrac{3}{2}^+\ $ & $N(1900)\tfrac{3}{2}^+\ $ \\\hline
$1^+\ $ & $70.1\ $ & $58.7\ $                 & $50.0\ $ & $71.7\ $ \\
$0^+\ $ & $20.3\ $ & $\phantom{9}7.4\ $ & $44.4\ $ & $\phantom{9}0.0\ $ \\
$0^-\ $ & $\phantom{9}6.2\ $ & $\phantom{9}3.4\ $ & $\phantom{9}3.6\ $ & $\phantom{9}0.0\ $ \\
$1^-\ $ & $\phantom{9}3.4\ $ & $30.5\ $ & $\phantom{9}2.0\ $ & $28.3\ $ \\\hline
\end{tabular*}

\end{center}
\end{table}

\begin{figure}[t]
\vspace*{2ex}

\leftline{\hspace*{0.5em}{\large{\textsf{A}}}}
\vspace*{-4ex}
\includegraphics[width=0.45\textwidth]{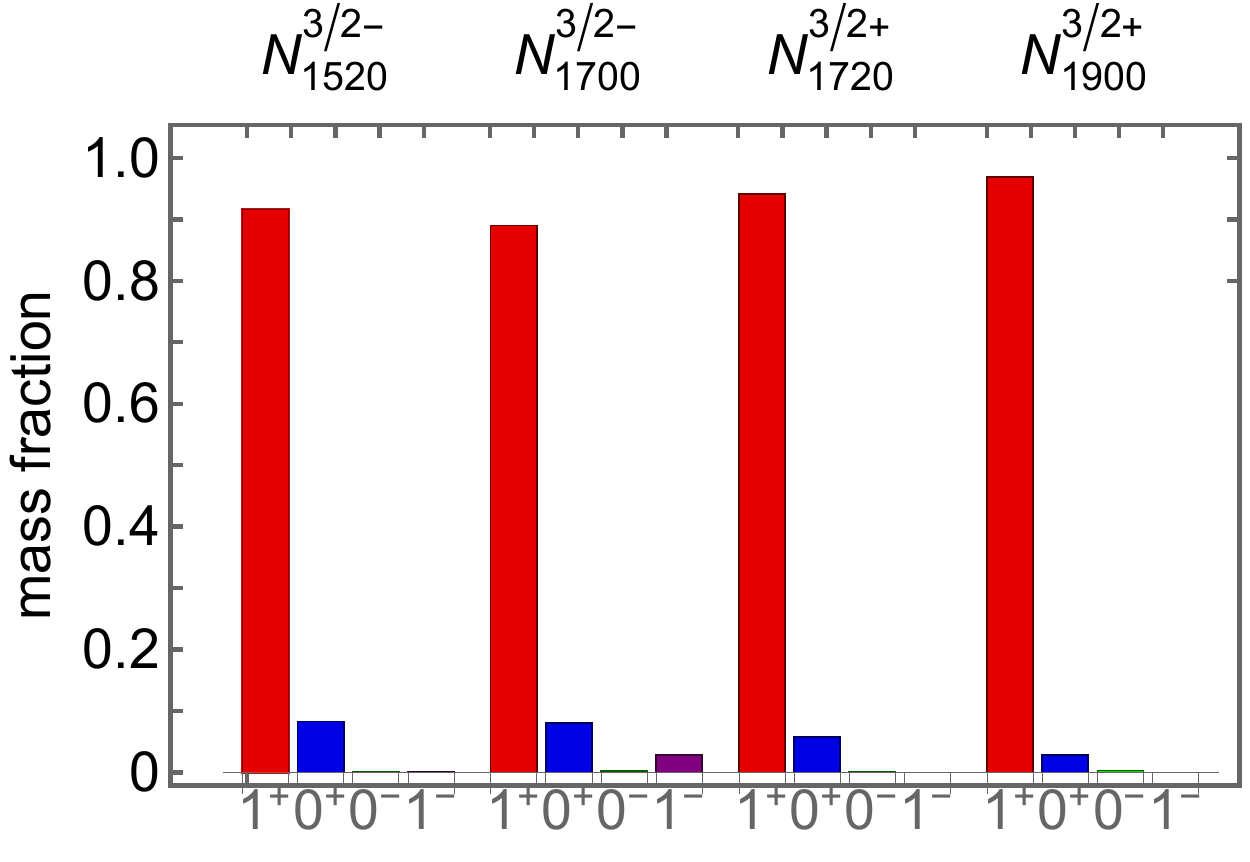}
\vspace*{1ex}

\leftline{\hspace*{0.5em}{\large{\textsf{B}}}}
\vspace*{-4ex}
\includegraphics[width=0.45\textwidth]{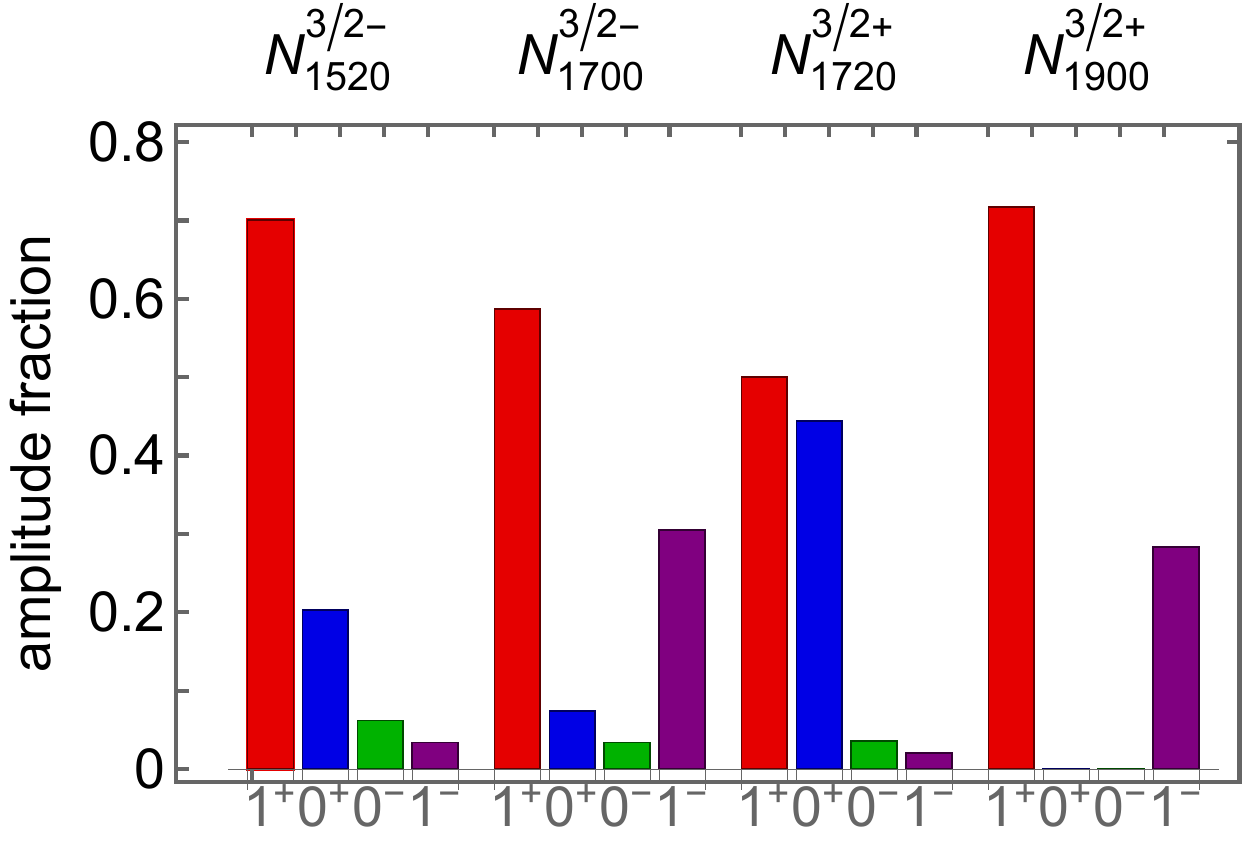}
\vspace*{-2ex}

\caption{\label{figureqqmass}
{\sf Panel A}.
Fractional contribution to mass of $(\tfrac{1}{2},\tfrac{3}{2}^P)$ baryons as additional diquark correlations are progressively included.
{\sf Panel B}.
Fraction of a given $(\tfrac{1}{2},\tfrac{3}{2}^P)$ baryon's Faddeev amplitude  contributed by distinct diquark correlations, defined in connection with Eq.\,\eqref{FractionFA}.
}
\end{figure}

Herein, for comparison with experiment, following Refs.\,\cite{Chen:2017pse, Yin:2021uom, Liu:2022ndb}, we subtract the mean value of the difference between our calculated masses and the real part of the related empirical pole-positions: $\delta^{N_{3/2}}_{\rm MB}=0.13\,$GeV.
The resulting comparison is displayed in Fig.\,\ref{MassCompare}.  The calculated level orderings and splittings match well with experiment, just as they do in analogous comparisons drawn from the results in Refs.\,\cite{Chen:2017pse, Liu:2022ndb}, which are also shown.
These predictions might be used to assist in refining dynamical coupled channels models by providing constraints on the size of meson-baryon final-state interactions in distinct $J^P$ channels.

In Table~\ref{qqmassfractions}A we list the mass obtained for each $(\tfrac{1}{2},\tfrac{3}{2}^\mp)$ system when the Faddeev equation is solved by keeping only one type of diquark correlation, then two, then three, and then all.  Evidently, in each case, once the axialvector and scalar diquarks are included, the mass is practically unchanged by including the other correlations.
This is highlighted by Table~\ref{qqmassfractions}B, which lists the change in a baryon's mass generated by the progressive inclusion of additional diquark correlations, in the order $1^+ \to 1^+, 0^+ \to 1^+, 0^+, 0^- \to 1^+, 0^+, 0^- 1^-$.  A pictorial representation of these results is provided in Fig.\,\ref{figureqqmass}.

\begin{table}[t]
\caption{\label{Lqdmassfractions}
{\sf Panel A}.
Calculated masses of the low-lying $(\tfrac{1}{2},\tfrac{3}{2}^\mp)$ baryons as obtained by stepwise including different orbital angular momentum components in the rest-frame Faddeev wave function.  The italicised entries indicate the lowest mass obtained for the given state when solving with a single partial wave.
{\sf Panel B}.
Change in the baryon's mass generated by progressive inclusion of additional orbital angular momentum components in its rest-frame Faddeev wave function.  This information is also depicted in Fig.\,\ref{Ldqmass}.  Naturally, the sum of entries in each column is unity.
(All masses in GeV.)
 }
\begin{center}
\begin{tabular*}
{\hsize}
{
l@{\extracolsep{0ptplus1fil}}
|c@{\extracolsep{0ptplus1fil}}
|c@{\extracolsep{0ptplus1fil}}
|c@{\extracolsep{0ptplus1fil}}
|c@{\extracolsep{0ptplus1fil}}}\hline
{\sf A}. mass\; &
$N(1520)\tfrac{3}{2}^-\ $ & $N(1700)\tfrac{3}{2}^-\ $ & $N(1720)\tfrac{3}{2}^+\ $ & $N(1900)\tfrac{3}{2}^+\ $ \\\hline
{\sf P} & $\mathit{1.70}\ $ & $\mathit{1.81}\ $  & $1.97\ $ & $2.11\ $ \\
{\sf S} & $2.02\ $ & $2.20\ $  & $1.88\ $ & $2.15\ $ \\
{\sf D} & $2.04\ $ & $2.14\ $  & $\mathit{1.77}\ $ & $\mathit{2.05}\ $ \\
{\sf F} & $2.05\ $ & $2.28\ $  & $2.36\ $ & $2.50\ $ \\\hline
{\sf PS} & $1.70\ $ & $1.81\ $  & $2.12\ $ & $2.13\ $ \\
{\sf PD} & $1.67\ $ & $1.81\ $  & $1.80\ $ & $1.97\ $ \\
{\sf SD} & $2.14\ $ & $2.18\ $  & $1.77\ $ & $1.89\ $ \\\hline
{\sf PSD} & $1.68\ $ & $1.81\ $  & $1.80\ $ & $2.05\ $ \\\hline
{\sf PSDF} & $1.68\ $ & $1.82\ $  & $1.78\ $ & $2.05\ $ \\\hline
\end{tabular*}

\medskip

\begin{tabular*}
{\hsize}
{
l@{\extracolsep{0ptplus1fil}}
|c@{\extracolsep{0ptplus1fil}}
|c@{\extracolsep{0ptplus1fil}}
|c@{\extracolsep{0ptplus1fil}}
|c@{\extracolsep{0ptplus1fil}}}\hline
{\sf B}. mass \%\; &
$N(1520)\tfrac{3}{2}^-\ $ & $N(1700)\tfrac{3}{2}^-\ $ & $N(1720)\tfrac{3}{2}^+\ $ & $N(1900)\tfrac{3}{2}^+\ $ \\\hline
{\sf P} & ${98.4}\ $ & ${99.5}\ $  & $\phantom{9}1.8\ $ & $\phantom{9}3.7\ $ \\
{\sf S} & $\phantom{9}0.1\ $ & $\phantom{9}0.4\ $  & $\phantom{9}0.2\ $ & $\phantom{9}3.6\ $ \\
{\sf D} & $\phantom{9}1.5\ $ & $\phantom{9}0.1\ $  & $97.2\ $ & $92.7\ $ \\
{\sf F} & $\phantom{9}0.0\ $ & $\phantom{9}0.0\ $  & $\phantom{9}0.8\ $ & $\phantom{9}0.0\ $ \\\hline
\end{tabular*}

\end{center}
\end{table}

A broadly consistent yet slightly different picture appears when one considers baryon Faddeev amplitudes.  Defining
\begin{equation}
\label{nAdefine}
{\mathpzc n}_{\mathpzc A}
= \sum_{j=+,0}\sum_{i=1}^8 \int \dfrac{d^4k}{(2\pi)^4} |{\mathpzc v}_{1^+}^{ji}(k^2, k \cdot Q)|^2 \,,
\end{equation}
with analogous expressions for ${\mathpzc n}_{\mathpzc S}$, ${\mathpzc n}_{\mathpzc P}$, ${\mathpzc n}_{\mathpzc V}$, then the ratios
\begin{equation}
\label{FractionFA}
{\mathbb F}_{\mathpzc C} = {\mathpzc n}_{\mathpzc C}/{\mathpzc n}_{\mathpzc T}\,,
\end{equation}
where ${\mathpzc C}= {\mathpzc S}$, ${\mathpzc A}$, ${\mathpzc P}$, ${\mathpzc V}$ and ${\mathpzc n}_{\mathpzc T} = {\mathpzc n}_{\mathpzc S} +{\mathpzc n}_{\mathpzc A} +{\mathpzc n}_{\mathpzc P} +{\mathpzc n}_{\mathpzc V}$, provide an indication of the relative strength of each diquark correlation in the baryon Faddeev amplitude.
The calculated results are listed in Table~\ref{qqmassfractions}C and drawn in Fig.\,\ref{figureqqmass}.
Axialvector diquarks dominate the Faddeev amplitude in all cases.  This is partly because each baryon contains two axialvector isospin projections, whereas all other diquarks are isoscalar, and axialvector diquarks have eight distinct contributing spinor structures.  Nevertheless, channel dynamics is playing a role because the same statements are true for the proton and yet the proton amplitude is dominated by the scalar diquark \cite[Fig.\,2]{Liu:2022ndb}.  The scalar diquark is also prominent in $N(1720)\tfrac{3}{2}^+$, which is the parity partner of the $N(1520)\tfrac{3}{2}^-$; so, the differences between their amplitude fractions owe to EHM.  It is interesting that the amplitudes of the other two parity partners, \emph{viz}.\ $N(1700)\tfrac{3}{2}^-$, $N(1900)\tfrac{3}{2}^+$, both contain visible vector diquark fractions, when defined according to Eq.\,\eqref{FractionFA}, and especially notable that the latter contains practically no scalar or pseudoscalar diquarks.
%

Such structural features can be tested in measurements of resonance electroexcitation at large momentum transfers \cite{Brodsky:2020vco, Carman:2020qmb, Mokeev:2021dab, Mokeev:2022xfo}.
In this connection, $N(1520)\tfrac{3}{2}^-$ electrocouplings are already available on the momentum transfer domain $Q_\gamma^2 \lesssim 5.5m_p^2$ \cite{CLAS:2009ces, Mokeev:2015lda, Mokeev:2020vab, Carman:2020qmb, Mokeev:2021dab, Mokeev:2022xfo}, where $m_p$ is the proton mass.  Regarding the other states mentioned here, extraction of electrocouplings on $Q_\gamma^2 \lesssim 5.5m_p^2$ is underway and results are expected within two years \cite{Mokeev:2020vab}.  Future experiments will collect data that can be used to determine the electrocouplings of most nucleon resonances out to $Q_\gamma^2 \approx 10 m_p^2$ \cite{Brodsky:2020vco, Mokeev:2021dab}.  This information should greatly assist in revealing measurable expressions of EHM \cite{Roberts:2020hiw, Roberts:2021xnz, Roberts:2021nhw, Binosi:2022djx, Papavassiliou:2022wrb}.

\section{Orbital Angular Momentum}
\label{SecAngular}
Given quark model expectations for $(\tfrac{1}{2},\tfrac{3}{2}^\mp)$ baryons, sketched in Sec.\,\ref{SecIntro}\;notes\,(\ref{N1520})\,--\,(\ref{N1900}), it is especially interesting to consider the quark+diquark baryon rest frame orbital angular momentum, $L_{qd}$, decomposition obtained from their Poincar\'e-covariant wave functions.

One means by which to measure the strength of the various $L_{qd}$ components is to solve the Faddeev equation for the wave function in the rest frame with first only one orbital angular momentum component and then steadily increase the $L_{qd}$ complexity:
(\emph{i}) ${\mathsf P}$-wave only; (\emph{ii}) ${\mathsf S}$-wave only; (\emph{iii}) ${\mathsf D}$-wave only; (\emph{iv}) ${\mathsf F}$-wave only; (\emph{v}) ${\mathsf P}+{\mathsf S}$-wave only; etc.  The results are presented in Table~\ref{Lqdmassfractions}A and depicted in Fig.\,\ref{Ldqmass}.

\begin{figure}[t]
\centerline{%
\includegraphics[clip, width=0.45\textwidth]{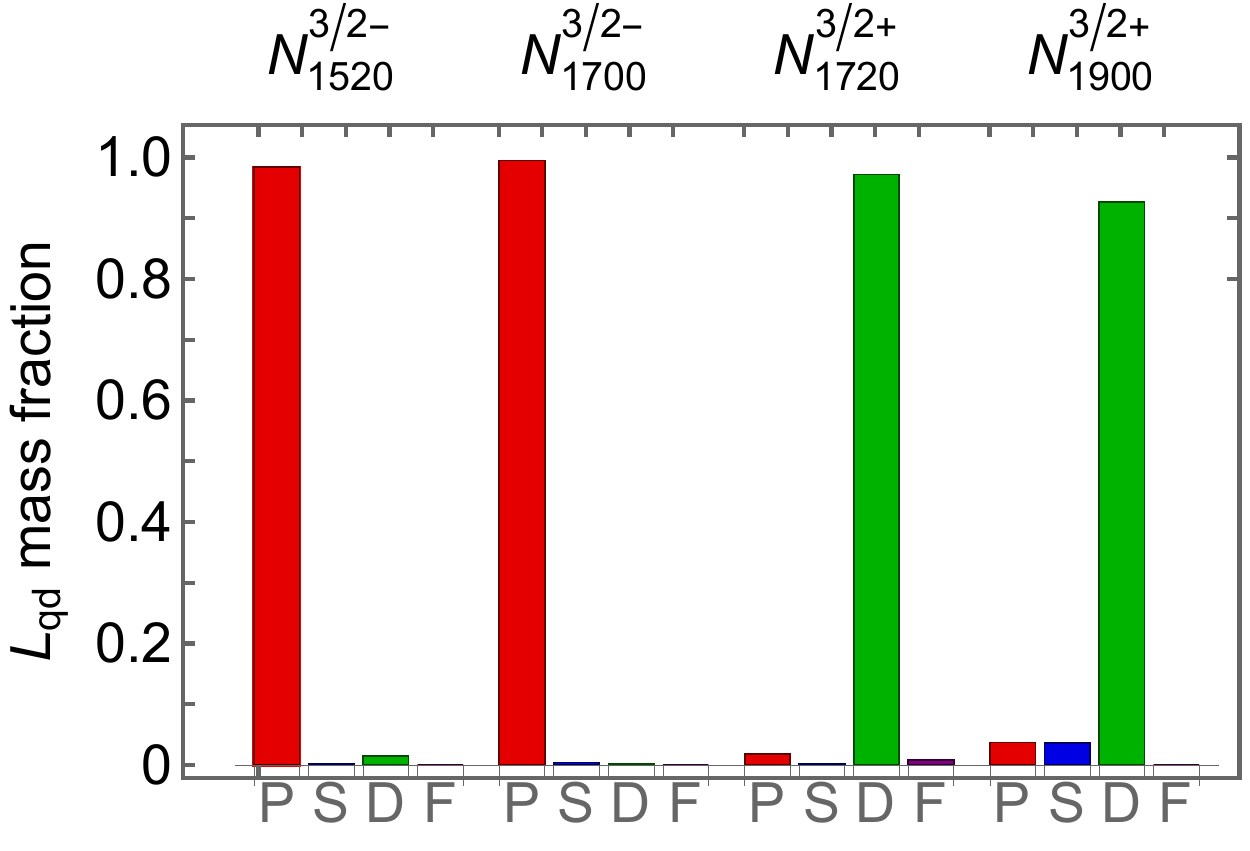}}
\caption{\label{Ldqmass}
Pictorial representation of Table~\ref{Lqdmassfractions}B.
Mass fraction contribution from each rest frame partial wave in the baryon wave function, calculated as follows:
$N(1520)\tfrac{3}{2}^-$, $N(1700)\tfrac{3}{2}^-$ -- begin with ${\mathsf P}$, then add ${\mathsf D}$, ${\mathsf S}$, ${\mathsf F}$;
and
$N(1900)\tfrac{3}{2}^+$, $N(1900)\tfrac{3}{2}^+$ -- begin with ${\mathsf D}$, then add ${\mathsf P}$, ${\mathsf S}$, ${\mathsf F}$.
}
\end{figure}

\begin{figure*}[!t]
\hspace*{-1ex}\begin{tabular}{lcl}
\large{\textsf{A}} & & \large{\textsf{B}}\\[-0.7ex]
%
\includegraphics[clip, width=0.41\textwidth]{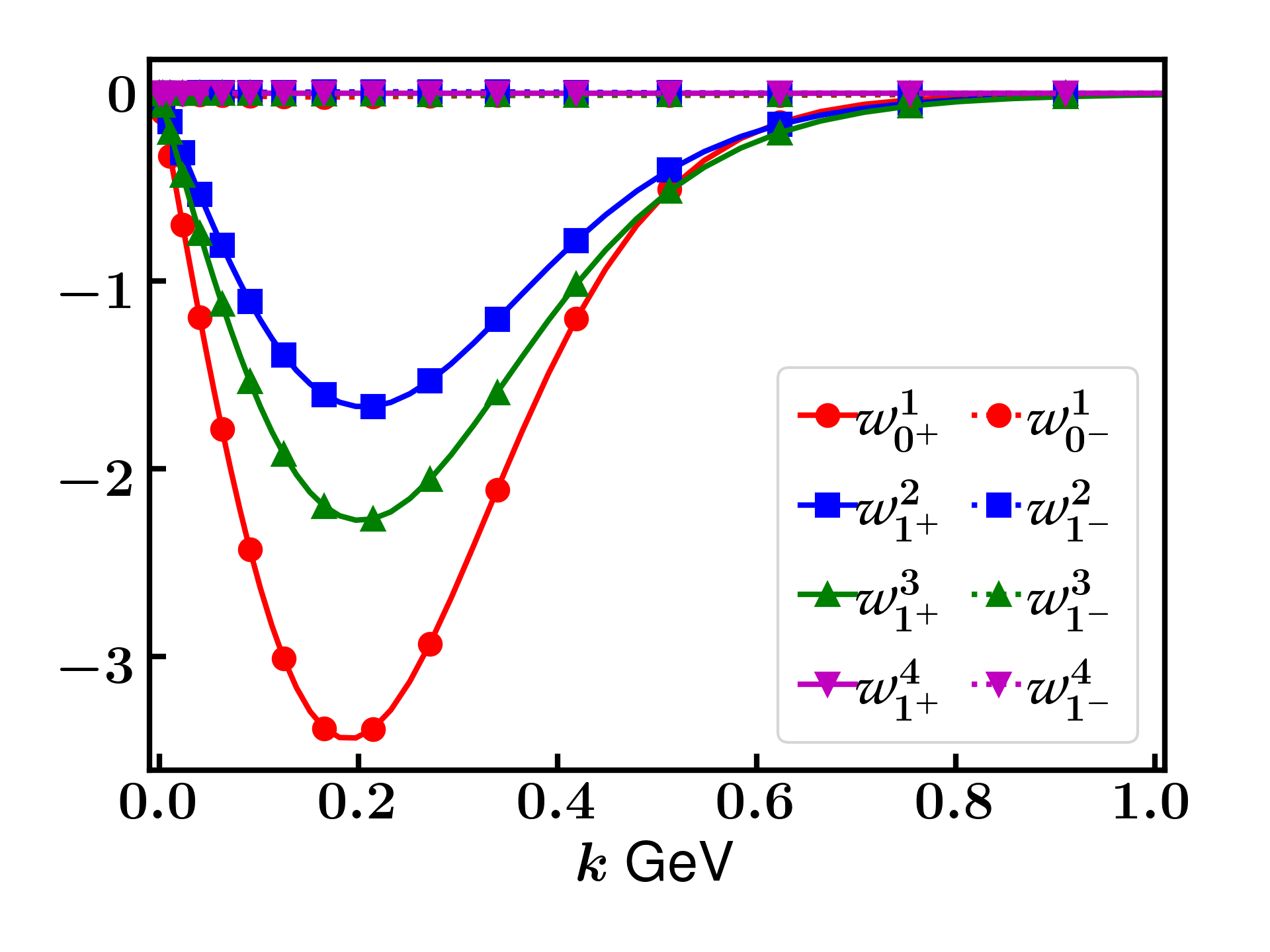} & \hspace*{4em} &
\includegraphics[clip, width=0.41\textwidth]{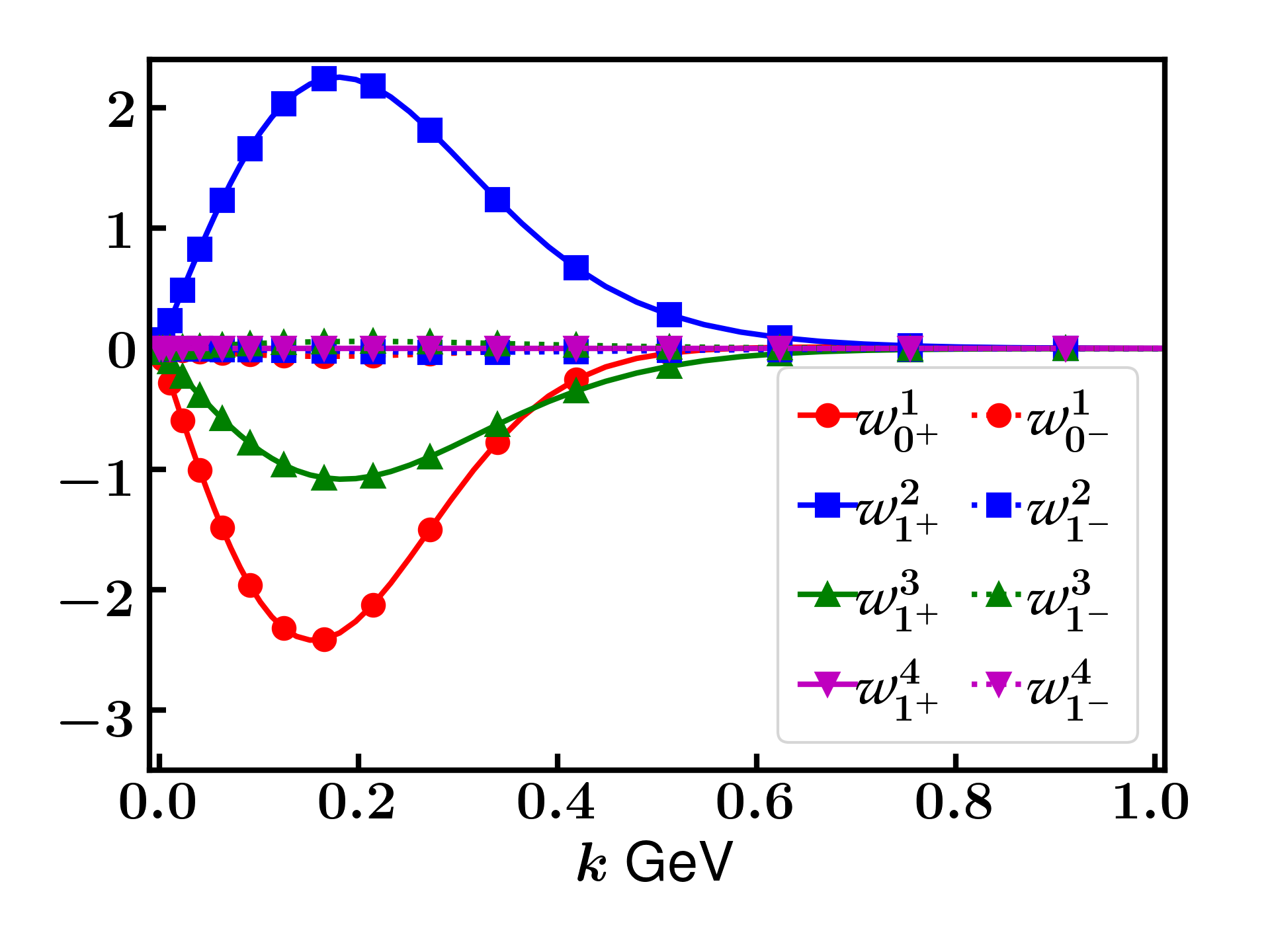} \\
\large{\textsf{C}} & & \large{\textsf{D}}\\[-0.7ex]
\includegraphics[clip, width=0.41\textwidth]{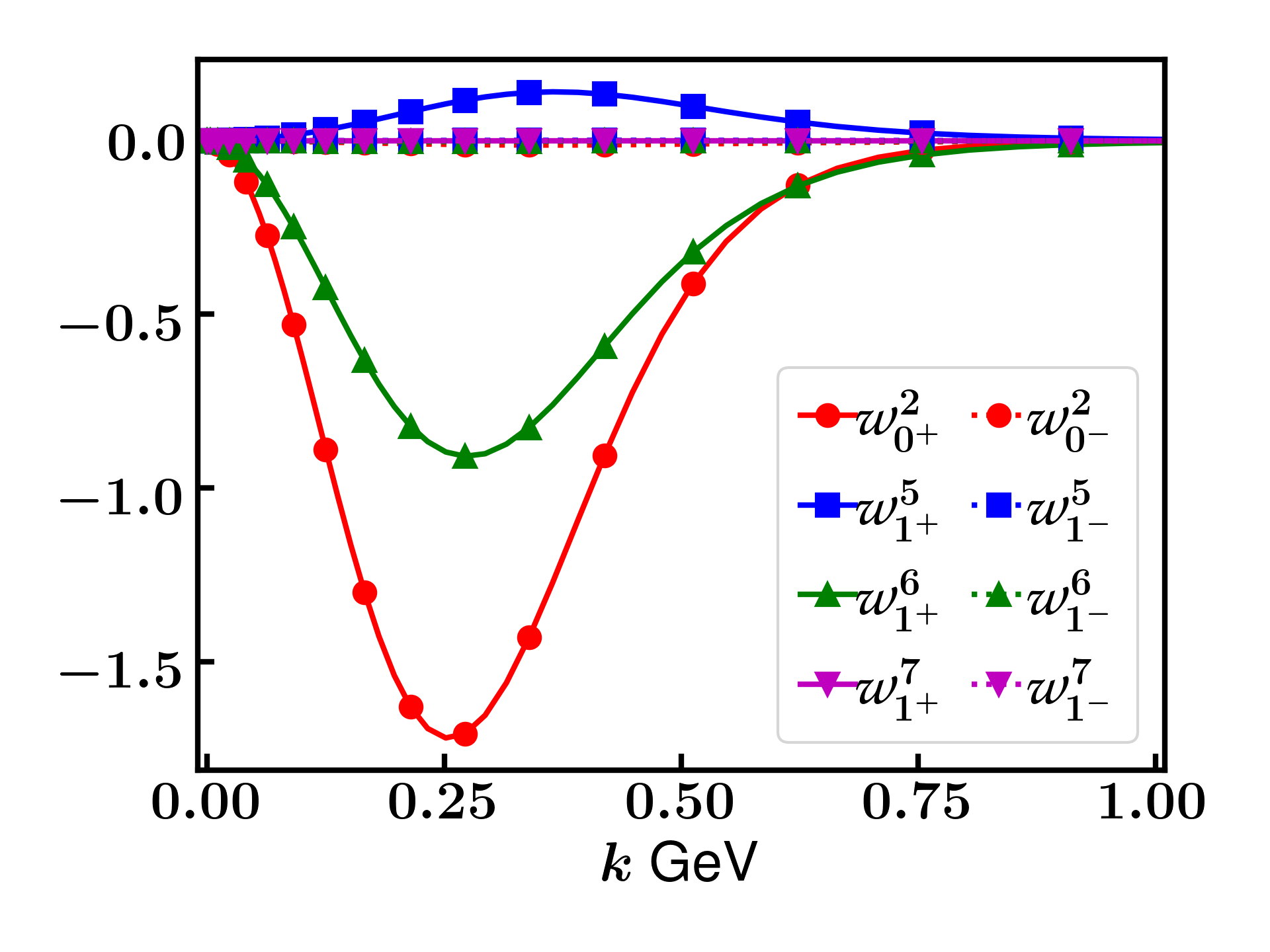} & \hspace*{4em} &
\includegraphics[clip, width=0.41\textwidth]{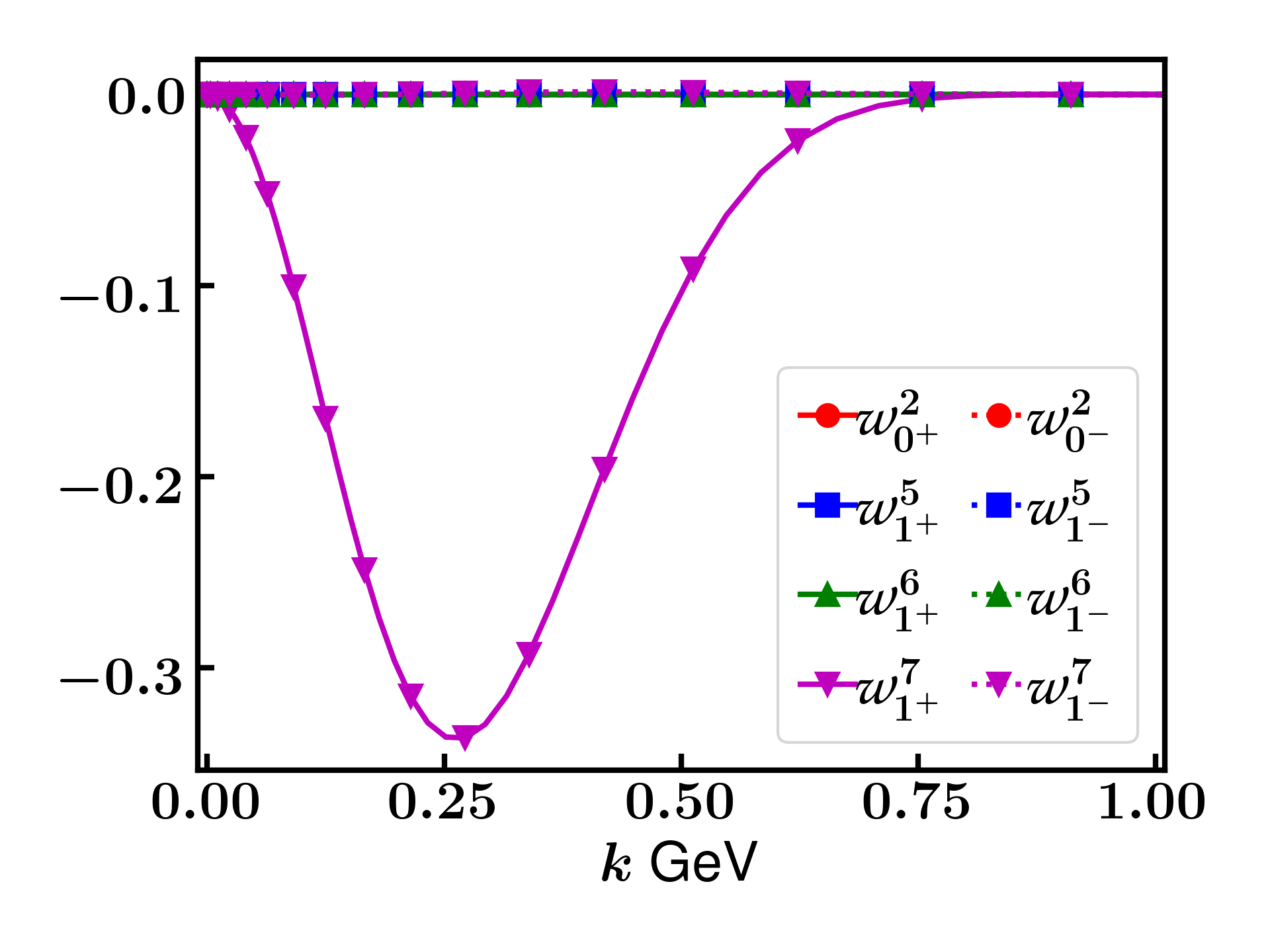} \\
\end{tabular}
\caption{\label{FChebyshev}
Zeroth Chebyshev moments -- Eq.\,\eqref{Chebyshev}.
\emph{Upper panels}.  Rest-frame $\mathsf P$-wave components in wave functions of the negative parity baryons: {\sf A} -- $N(1520)\tfrac{3}{2}^-$; and {\sf B} -- $N(1700)\tfrac{3}{2}^-$.
\emph{Lower panels}.  Rest-frame $\mathsf D$-wave components in wave functions of the positive parity baryons: {\sf C} -- $N(1900)\tfrac{3}{2}^+$; and {\sf D} -- $N(1900)\tfrac{3}{2}^+$.}
\end{figure*}

It is worth highlighting some insights revealed by Table~\ref{Lqdmassfractions}.
\begin{enumerate}[(i)]
\item For each state, a solution is obtained using only one partial wave -- $\mathsf P$, $\mathsf D$, $\mathsf S$, or $\mathsf F$, and with any subset of the complete array of partial waves.  Evidently, the quark-exchange kernel in Fig.\,\ref{FigFaddeev} is very effective at binding $(\tfrac{1}{2},\tfrac{3}{2}^\mp)$ baryons, just as it is in every other channel considered hitherto \cite{Chen:2017pse, Liu:2022ndb}.
\item Considering only single partial waves, then that which produces the lowest mass should serve as a good indicator of the dominant orbital angular momentum component in the state.  This gross measure leads to the following assignments.
    $N(1520)\tfrac{3}{2}^-$ and $N(1700)\tfrac{3}{2}^-$ are largely ${\mathsf P}$ wave in character; and
    $N(1720)\tfrac{3}{2}^+$ and $N(1900)\tfrac{3}{2}^+$ are largely ${\mathsf D}$ wave states.
Consequently, drawn with this broad-brush, the orbital angular momentum structure of each $(\tfrac{1}{2},\tfrac{3}{2}^+)$ baryon matches that which is typical of quark models, so long as the orbital angular momentum is identified with that of a quark+diquark system.

Notwithstanding these remarks, a more complicated structural picture will be revealed below.
\end{enumerate}

The link with quark models may be augmented by considering the zeroth Chebyshev projection of the dominant component of a baryon's Faddeev wave function, as measured by the italicised entries in Table~\ref{Lqdmassfractions}A, \emph{viz}.
\begin{equation}
\label{Chebyshev}
{\mathpzc w}^j(k^2) = \frac{2}{\pi}\int_{-1}^1 dx\,\sqrt{1-x^2}\,{\mathpzc w}^j(k^2,x \sqrt{k^2 Q^2})\,,
\end{equation}
with the terms identified using Eqs.\,\eqref{ScalarFunctions} and Table~\ref{TabL} of the appendix.  The results, drawn in Fig.\,\ref{FChebyshev}, highlight that these amplitudes do not exhibit an obvious zero.  Looking at the same projection of the subdominant partial waves, one finds that only a few possess such a zero.  Consequently, it may reasonably be concluded that no state should be considered a radial excitation of another; hence, the collection of $(\tfrac{1}{2},\tfrac{3}{2}^\mp)$ baryons form a set of states related via $L_{qd}$ excitation.
Such structural predictions, too, can be tested via comparisons with data obtained on the $Q_\gamma^2$-dependence of nucleon-to-resonance transition form factors \cite{Brodsky:2020vco, Carman:2020qmb, Mokeev:2021dab, Mokeev:2022xfo}.

\begin{figure}[!t]
\vspace*{1ex}

\centerline{%
\includegraphics[clip, width=0.45\textwidth]{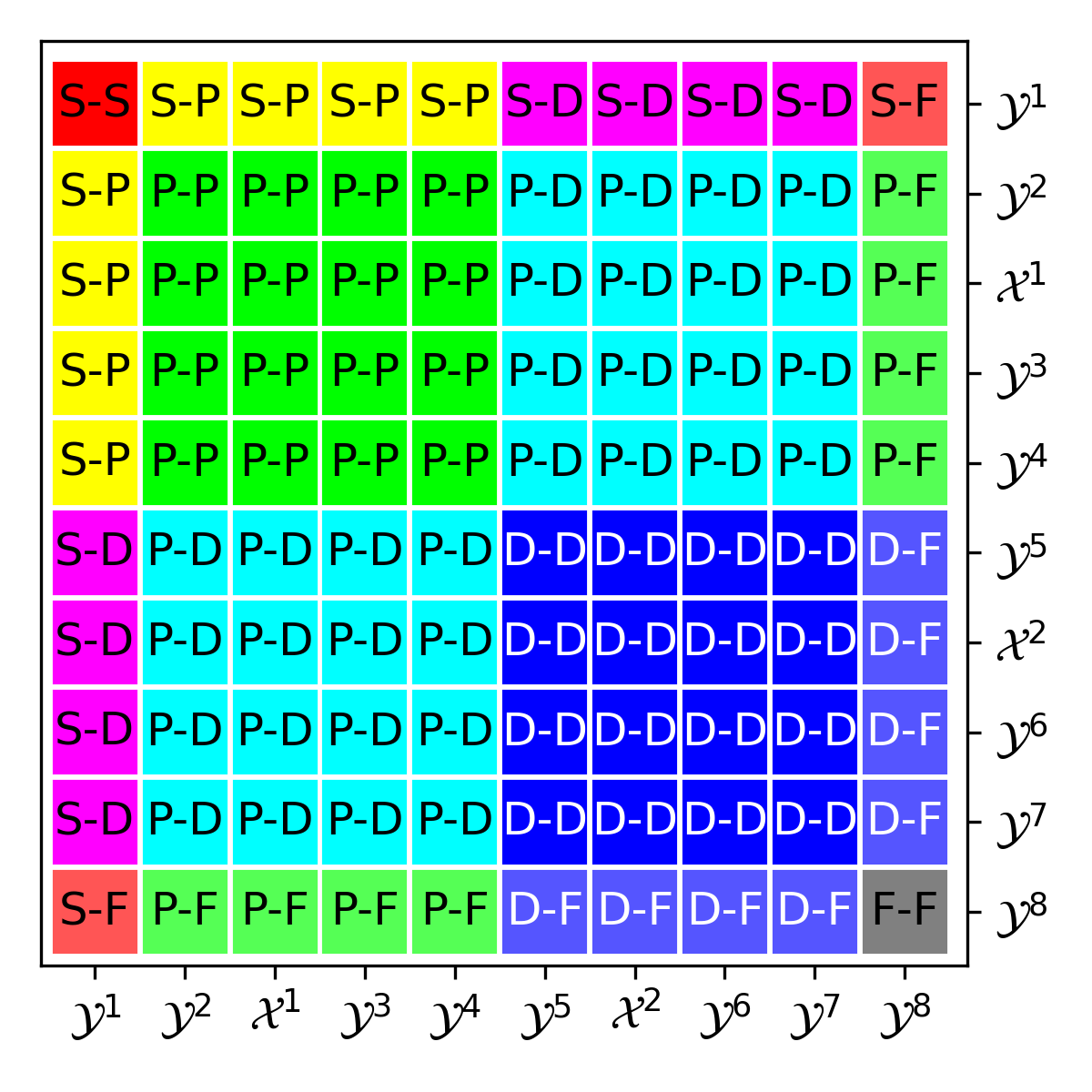}}
\caption{\label{LWFlegend}
Legend for interpretation of Figs.\,\ref{LFigures}A\,--\,D, indicating both direct terms and interference overlaps between the various identified orbital angular momentum basis components in the baryon rest frame.
}
\end{figure}

\begin{figure*}[!t]
\hspace*{-1ex}\begin{tabular}{lcl}
\large{\textsf{A}} & & \large{\textsf{B}}\\[-0.7ex]
%
\includegraphics[clip, width=0.43\textwidth]{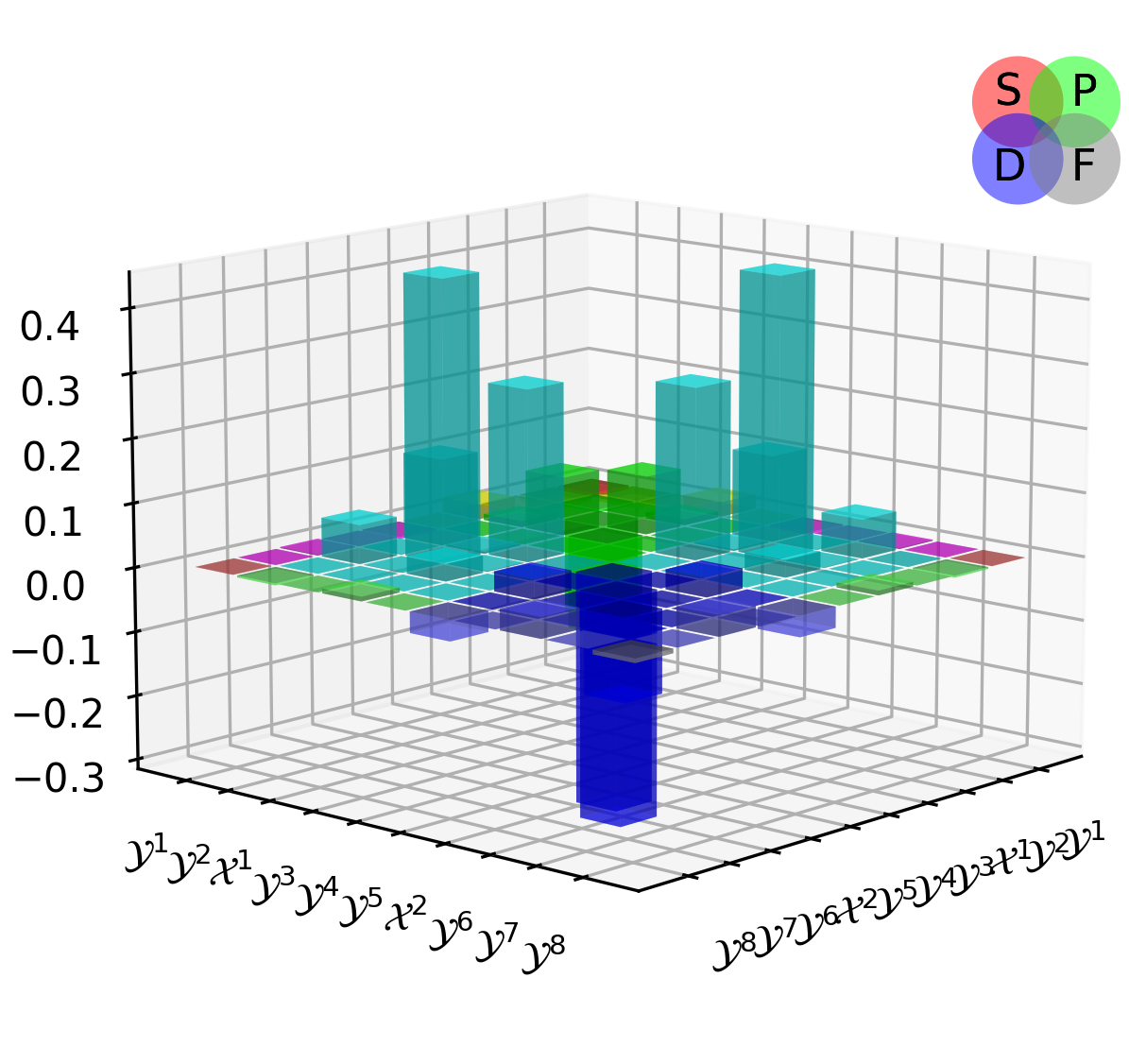} & \hspace*{4em} &
\includegraphics[clip, width=0.43\textwidth]{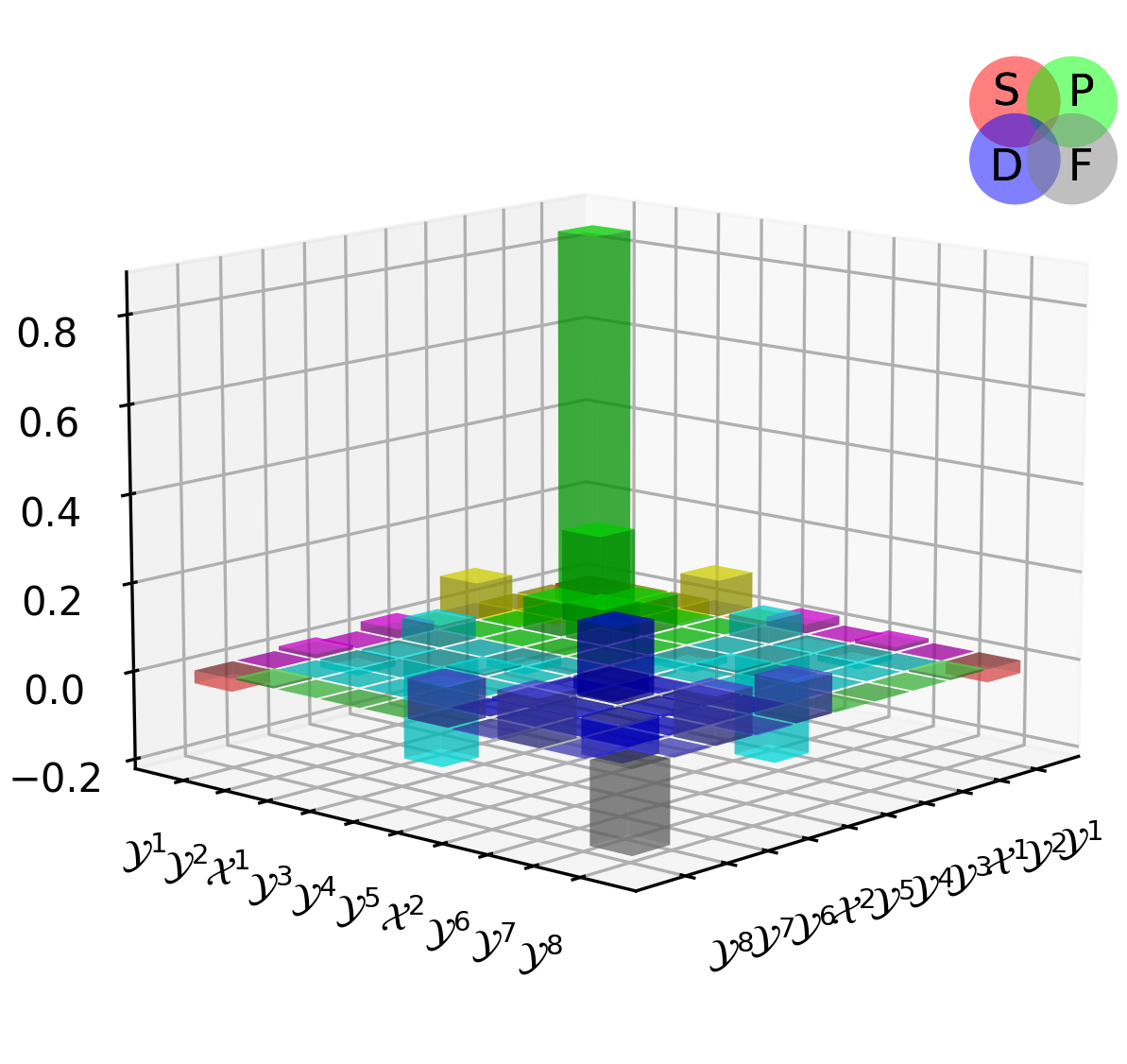} \\
\large{\textsf{C}} & & \large{\textsf{D}}\\[-0.7ex]
\includegraphics[clip, width=0.43\textwidth]{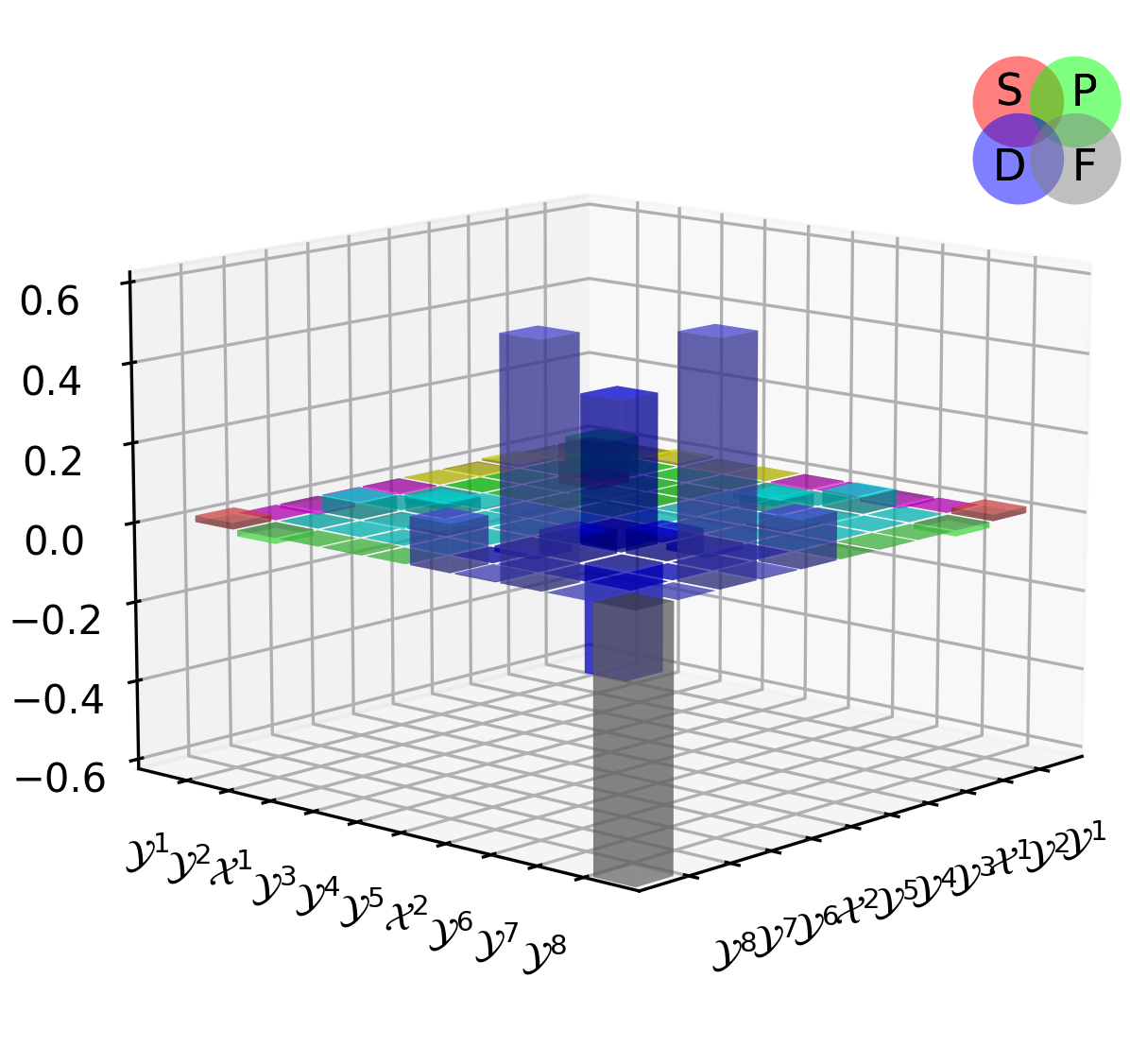} & \hspace*{4em} &
\includegraphics[clip, width=0.43\textwidth]{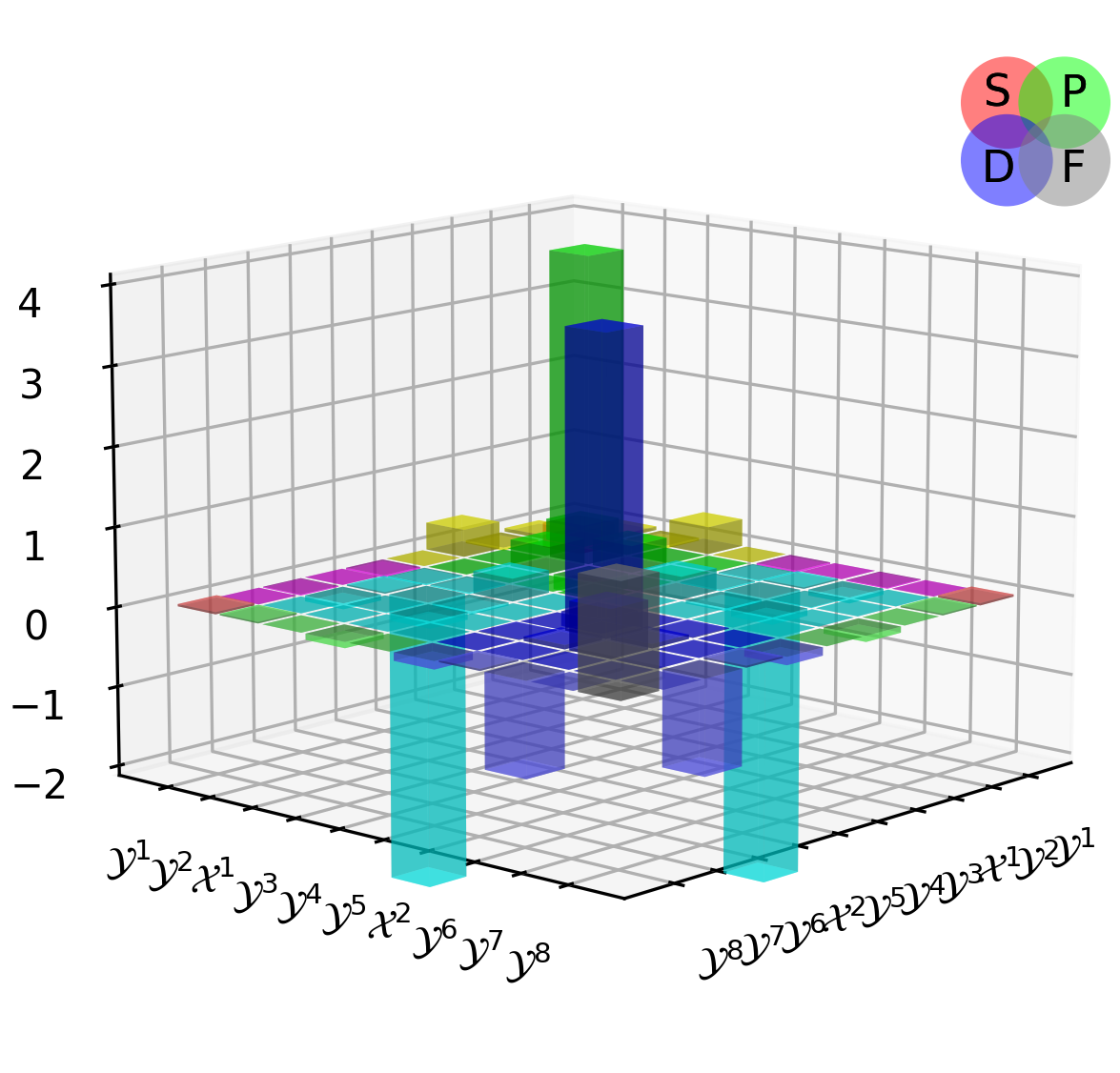} \\
\end{tabular}
\caption{\label{LFigures}
Rest frame quark+diquark orbital angular momentum content of $(\tfrac{1}{2},\tfrac{3}{2}^\mp)$ states considered herein, as measured by the contribution of the various components to the associated canonical normalisation constant:
{\sf A} -- $N(1520)\tfrac{3}{2}^-$;
{\sf B} -- $N(1700)\tfrac{3}{2}^-$;
{\sf C} -- $N(1720)\tfrac{3}{2}^+$; and
{\sf D} -- $N(1900)\tfrac{3}{2}^+$.
The images are drawn according to the legend in Fig.\,\ref{LWFlegend} and with reference to Table~\ref{TabL}, the basis in Eqs.\,\eqref{XYbasis}, and the expansion in Eqs.\,\eqref{ScalarFunctions}.  Only scalar and axialvector components are retained because they contribute most to each baryon's mass.
There are both positive (above plane) and negative (below plane) contributions to the overall normalisations, which are all positive.}
\end{figure*}

It has often been highlighted that masses are long-wavelength observables, whose values are not very sensitive to the finer structural details expressed in a baryon's wave function \cite{Roberts:2021xnz, Liu:2022ndb}.  The apparent simplicity of the results in Fig.\,\ref{Ldqmass} is thus somewhat misleading.  This is exposed, \emph{e.g}., by performing an $L_{qd}$ breakdown of each baryon's canonical normalisation, a quantity that is related to the zero momentum transfer value of the electric form factor of the valence quarks within the state; hence, observable.\footnote{Expressed using the wave function, the canonical normalisation integrand is a sum of terms, each of which involves an inverse diquark propagator.  Such functions exhibit singularities that are cancelled during integration by zeros in the wave functions.  Evaluating the integrals numerically, one must use an algorithm that ensures such cancellations are perfect.  No such issues arise when evaluating the normalisation using the Faddeev amplitude.}
Working with the assignments identified in Fig.\,\ref{LWFlegend}, those decompositions are depicted in Fig.\,\ref{LFigures}.  These figures are drawn from the tables collected in Appendix~\ref{AAngular}.  Since negative-parity diquarks make negligible contributions to a baryon's mass -- Fig.\,\ref{figureqqmass}A, only the scalar and axialvector contributions are recorded.

Consider first Fig.\,\ref{LFigures}A, which displays the rest-frame $L_{qd}$-breakdown of the $N(1520)\tfrac{3}{2}^-$ canonical normalisation constant.  Evidently, the most prominent positive contributions are provided by constructive $\mathsf P\otimes \mathsf D$-wave interference terms; contributions from purely $\mathsf P$-wave components are visible, but interfere destructively; and pure $\mathsf D$-wave terms are responsible for largely destructive interference.  Whilst these observations are consistent with the results in Fig.\,\ref{Ldqmass}, they also reveal the structural complexity of a Poincar\'e covariant wave function.
This detailed picture is very different from that obtained from quark models built upon (weakly-broken) SU$(6)$ spin-flavour symmetry, as listed, \emph{e.g}., in Sec.\,\ref{SecIntro}\;note (\ref{N1520}).
Given that resonance electroexcitation data on this state are available out to momentum transfers $Q_\gamma^2 \approx 5.5 m_p^2$ \cite[Table~4]{Mokeev:2022xfo}, then our Faddeev equation structural predictions can be tested once the wave functions are used to calculate the associated transition form factors.
Where such comparisons have already been made, the Faddeev equation predictions have been validated \cite{Burkert:2019bhp, Roberts:2018hpf, Lu:2019bjs, VictorMokeevPresentation}.

Turning to Fig.\,\ref{LFigures}B, the $N(1700)\tfrac{3}{2}^-$ wave function is seen to be less complex than that of the $N(1520)\tfrac{3}{2}^-$.  True to Fig.\,\ref{Ldqmass}, pure $\mathsf P$-wave contributions to the canonical normalisation are dominant; there is some destructive $\mathsf P\otimes \mathsf D$-wave interference; simple $\mathsf D$-wave contributions largely cancel amongst themselves; and $\mathsf D\otimes \mathsf F$-wave constructive interference offsets a destructive $\mathsf F$-wave contribution.
In this case, resonance electroexcitation data is available for $Q_\gamma^2 \lesssim 1.7 m_p^2$ \cite{Mokeev:2020hhu}. However, data at larger $Q_\gamma^2$ would be needed to test our structural predictions.

Drawn in Fig.\,\ref{LFigures}C, the $N(1720)\tfrac{3}{2}^+$ wave function is simpler still.  This state is the parity partner of the $N(1520)\tfrac{3}{2}^-$, so differences between their wave functions are driven by EHM.  Consistent with Fig.\,\ref{Ldqmass}, normalisation contributions related to $\mathsf  D$-waves are most prominent: the largest positive terms are generated by constructive $\mathsf D\otimes \mathsf F$-wave interference.  As in the $N(1700)\tfrac{3}{2}^-$, pure $\mathsf D$-wave contributions largely cancel amongst themselves; and there is a sizeable destructive $\mathsf F$-wave contribution.  Here, too, resonance electroexcitation data is only available for $Q_\gamma^2 \lesssim 1.7 m_p^2$ \cite{Mokeev:2020hhu}.
Our quark+diquark Faddeev equation does not generate the $N^\prime(1720)\tfrac{3}{2}^+$ state discussed in Ref.\,\cite{Mokeev:2020hhu}.  It may conceivably appear if the diquark correlations were to possess a richer structure than described by the simplified forms in Eqs.\,\eqref{qqBSAs}.

The $N(1900)\tfrac{3}{2}^+$ normalisation strengths are displayed in Fig.\,\ref{LFigures}D.  This state is the parity partner of the $N(1700)\tfrac{3}{2}^-$.  Here, EHM is seen to drive very strong pure $\mathsf  P$- and $\mathsf D$-wave contributions.  There is also a prominent constructive $\mathsf F$-wave contribution; and $\mathsf P\otimes \mathsf D$-wave and $\mathsf D\otimes \mathsf F$-wave interference are strongly destructive.  Whilst consistent with the results drawn for this state in Fig.\,\ref{Ldqmass}, the detailed picture is again far more complex.
It is worth comparing the magnitude scale in Fig.\,\ref{LFigures}D with that of the other panels.  Owing to strong interference between partial waves, the $N(1900)\tfrac{3}{2}^+$ normalisation constant is roughly twice that determined in the other cases.  This further emphasises the complexity of its Poincar\'e-covariant wave function. %
There is currently no $N(1900)\tfrac{3}{2}^+$ resonance electroexcitation data \cite[Table~4]{Mokeev:2022xfo}.

\section{Summary and Perspective}
\label{epilogue}
Extending Refs.\,\cite{Chen:2017pse, Liu:2022ndb}, we employed a Poincar\'e-covariant Faddeev equation [Fig.\,\ref{FigFaddeev}] to deliver predictions for the masses and wave functions of the four lowest lying $(I,J^P) = (\tfrac{1}{2},\tfrac{3}{2}^\mp)$ baryons.  The Faddeev kernel is constructed using dressed-quark and nonpointlike diquark degrees-of-freedom and expresses two binding mechanisms: one is that within the diquark correlations themselves; and the other is generated by exchange of a dressed-quark, which emerges as one fully-interacting diquark splits up and is subsequently absorbed into formation of another.
This quark+diquark picture was introduced more than thirty years ago \cite{Cahill:1988dx, Burden:1988dt, Reinhardt:1989rw, Efimov:1990uz} and has since evolved into an efficacious tool for the prediction and explanation of baryon properties, including
the large-$Q_\gamma^2$ behaviour of elastic and transition form factors \cite{Cui:2020rmu, Chen:2018nsg},
axial form factors \cite{Chen:2020wuq, Chen:2021guo, ChenChen:2022qpy},
and parton distribution functions \cite{Chang:2022jri, Lu:2022cjx},

General considerations reveal that $(\tfrac{1}{2},\tfrac{3}{2}^\mp)$ baryons may contain five distinct types of diquark correlation: $(0,0^+)$, $(1,1^+)$, $(0,0^-)$, $(0,1^-)$, $(1,1^-)$; but our calculations showed that a good approximation is obtained by keeping only $(0,0^+)$, $(1,1^+)$ correlations [Sec.\,\ref{Solutions}].   This is not true for $(\tfrac{1}{2},\tfrac{1}{2}^-)$ states, in which $(0,0^-)$, $(0,1^-)$ diquarks are important \cite{Chen:2017pse, Raya:2021pyr}.

Exploiting our Poincar\'e-covariant Faddeev wave functions for $(\tfrac{1}{2},\tfrac{3}{2}^\mp)$ baryons, we drew connections and contrasts with structural expectations deriving from quark models built upon (weakly-broken) SU$(6)$ spin-flavour symmetry [Sec.\,\ref{SecAngular}].
In this context, the orbital angular momentum composition was of particular interest.  However, since the $J=L+S$ separation of total angular momentum into a sum of orbital angular momentum and spin is frame dependent and quark models typically express only Galilean covariance, we worked with rest-frame projections of our Poincar\'e-covariant Faddeev wave functions.
Viewed with low resolution and identifying orbital angular momentum as that which exists between dressed-quarks and -diquarks, $L_{qd}$, we found broad agreement.  Namely, the collection of $(\tfrac{1}{2},\tfrac{3}{2}^\mp)$ baryons form a set of states related via $L_{qd}$ excitation: the negative parity states are primarily $\mathsf P$-wave in nature whereas the positive parity states are $\mathsf D$ wave.

On the other hand, we also probed the structure of $(\tfrac{1}{2},\tfrac{3}{2}^\mp)$ baryons with finer resolution, using maps of the contributions to the canonical normalisation constants from the various $L_{qd}$ components of the Poincar\'e-covariant wave functions.  This revealed far greater complexity than typical of quark model descriptions [Fig.\,\ref{LFigures}], with significant interference between $L_{qd}$ components.
These structural predictions can be tested in comparisons between measurements of resonance electroexcitation at large momentum transfers and predictions for the associated resonance electroproduction form factors based on our wave functions.  Concerning the $N(1520)\tfrac{3}{2}^-$, data already exists that could be used for this purpose and the calculations are underway.   No such data exists for the other states and so our predictions are likely to encourage new experimental efforts in this area.

It is worth reiterating that the interpolating fields for negative and positive baryons are related by chiral rotation of the quark spinors used in their construction.  This entails that all differences between parity partner states owe fundamentally to chiral symmetry breaking, which is overwhelmingly dynamical in the light-quark sector \cite{Lane:1974he, Politzer:1976tv, Pagels:1979hd, Cahill:1985mh, Bashir:2012fs}.  Parity partner channels are identical when chiral symmetry is restored \cite{Roberts:2000aa, Fischer:2018sdj}.
Regarding the baryons considered herein, parity connects
$N(1520)\tfrac{3}{2}^-$--\,$N(1720)\tfrac{3}{2}^+$ and $N(1700)\tfrac{3}{2}^-$--\,$N(1900)\tfrac{3}{2}^+$;
and we have seen that, again like the $(\tfrac{1}{2},\tfrac{1}{2}^\pm)$ and $(\tfrac{3}{2},\tfrac{3}{2}^\pm)$ sectors, dynamical chiral symmetry breaking (DCSB) introduces marked differences between the internal structures of parity partners.  This has a marked influence on the mass splitting between the partner states and explains why it does not exhibit a simple pattern, \emph{viz}.\ empirically \cite{Workman:2022ynf}:
{\allowdisplaybreaks
\begin{equation}
\begin{array}{lc}
{\rm states} & {\rm mass\;splitting/GeV}\\
N(1535)\tfrac{1}{2}^- - N(940)\tfrac{1}{2}^+ & 0.57\,, \\
N(1650)\tfrac{1}{2}^- - N(1440)\tfrac{1}{2}^+& 0.29\,,\\
\Delta(1700)\tfrac{3}{2}^- - \Delta(1232)\tfrac{3}{2}^+ & 0.46 \,,\\
\Delta(1940)\tfrac{3}{2}^- - \Delta(1600)\tfrac{3}{2}^+ & 0.44 \,, \\
N(1720)\tfrac{3}{2}^+ - N(1520)\tfrac{3}{2}^- & 0.17 \,, \\
N(1900)\tfrac{3}{2}^+ - N(1700)\tfrac{3}{2}^- & 0.22 \,. \\
\end{array}
\end{equation}
}

DCSB is a corollary of emergent hadron mass (EHM); and confinement, too, may be argued to derive from EHM \cite{Roberts:2021nhw}.  Thus, validating our predictions of marked structural differences between parity partners throughout the hadron spectrum has the potential to reveal much of importance about the Standard Model.  As already noted, resonance electroexcitation experiments on $Q_\gamma^2 \gtrsim 2\,m_p^2$ are one way of achieving this goal.

Having completed this analysis, it would be natural to close the cycle and use the Poincar\'e-covariant quark+diquark Faddeev equation to develop structural insights into $(\tfrac{3}{2},\tfrac{1}{2}^\pm)$ baryons.
Further, in aiming to validate the pictures provided, it is essential to calculate the electromagnetic transition form factors for all states mentioned above.
Baryons containing heavier valence quarks may present additional opportunities, particularly because many models of such systems give special treatment to the heavier degrees-of-freedom, \emph{e.g}., Refs.\,\cite{Ebert:2011kk, Ferretti:2019zyh, Ebert:2002ig, Zhang:2008rt, Chen:2016spr, Li:2019ekr}, whereas our dynamical diquark picture indicates that all valence quarks should be treated equally \cite{Yin:2019bxe, Yin:2021uom}.
Furthermore, having highlighted the complexity of the Poincar\'e-covariant wave functions that describe quark-plus-dynamical-diquark systems,  then our results may also have implications for studies of the tetra- and penta-quark problems, which are typically treated using very simple pictures of diquark correlations and their interactions with other bound-state constituents \cite[Sec.\,3.6]{Barabanov:2020jvn}.

%
%
\begin{acknowledgments}
We are grateful for constructive comments from \mbox{Z.-F.~Cui}, Y.~Lu, L.~Meng, V.\,I.~Mokeev and J.~Segovia.
This work used the computer clusters at the Nanjing University Institute for Nonperturbative Physics.
Work supported by:
National Natural Science Foundation of China (grant nos.\,12135007, 12047502);
and
Jiangsu Province Fund for Postdoctoral Research (grant no.\,2021Z009).
\end{acknowledgments}

\newpage

\appendix

\begin{widetext}
\section{Quark+diquark angular momentum}
\label{AAngular}
Using our Faddeev equation solutions for the Poincar\'e-covariant baryon wave functions, evaluated in the rest frame, we computed the contributions of various quark+diquark orbital angular momentum components to each baryon's canonical normalisation constant.  The results are recorded in this appendix:
$N(1520)\tfrac{3}{2}^-$ -- Table~\ref{tab:L-decompz-n1520-sc-av-dq};
$N(1700)\tfrac{3}{2}^-$ -- Table~\ref{tab:L-decompz-n1700-sc-av-dq};
$N(1720)\tfrac{3}{2}^+$ -- Table~\ref{tab:L-decompz-n1720-sc-av-dq}; and
$N(1900)\tfrac{3}{2}^+$ -- Table~\ref{tab:L-decompz-n1900-sc-av-dq}.
The images in Fig.\,\ref{LFigures} are drawn from these tables.
\end{widetext}

	\begin{table}[th]
\caption{$N(1520)\tfrac{3}{2}^-$: quark+diquark orbital angular momentum breakdown of the canonical normalisation constant, drawn in Fig.\,\ref{LFigures}A.  The subarrays are composed according to the legend in Fig.\,\ref{LWFlegend} and the sum of all entries is unity. \label{tab:L-decompz-n1520-sc-av-dq}}
		\centering
		\begin{tabular}{c|c|cccc|cccc|c}
			\hline
			~ & $\mathcal{Y}^1$ & $ \mathcal{Y}^2 $ & $\mathcal{X}^1$ & $ \mathcal{Y}^3 $ & $ \mathcal{Y}^4 $ & $ \mathcal{Y}^5 $ & $ \mathcal{X}^2 $ & $ \mathcal{Y}^6 $ & $ \mathcal{Y}^7 $ & $ \mathcal{Y}^8 $ \\ \hline
			$\mathcal{Y}^1$ & -0.01 & -0.01 & 0.00 & 0.02 & 0.00 & 0.00 & 0.00 & 0.00 & 0.00 & 0.00  \\ \hline
			$ \mathcal{Y}^2 $ & -0.01 & -0.01 & 0.00 & 0.01 & 0.00 & 0.13 & 0.00 & 0.06 & 0.00 & 0.00  \\
			$\mathcal{X}^1$ & 0.00 & 0.00 & -0.15 & 0.09 & 0.00 & 0.00 & 0.44 & 0.00 & 0.00 & 0.00  \\
			$ \mathcal{Y}^3 $ & 0.02 & 0.01 & 0.09 & -0.15 & 0.00 & 0.27 & 0.00 & 0.03 & 0.00 & 0.01  \\
			$ \mathcal{Y}^4 $ & 0.00 & 0.00 & 0.00 & 0.00 & 0.00 & 0.00 & 0.00 & 0.00 & 0.00 & 0.00  \\ \hline
			$ \mathcal{Y}^5 $ & 0.00 & 0.13 & 0.00 & 0.27 & 0.00 & -0.36 & 0.00 & 0.03 & 0.00 & -0.03  \\
			$ \mathcal{X}^2 $ & 0.00 & 0.00 & 0.44 & 0.00 & 0.00 & 0.00 & -0.35 & 0.00 & 0.00 & 0.00  \\
			$ \mathcal{Y}^6 $ & 0.00 & 0.06 & 0.00 & 0.03 & 0.00 & 0.03 & 0.00 & -0.13 & 0.00 & 0.03  \\
			$ \mathcal{Y}^7 $ & 0.00 & 0.00 & 0.00 & 0.00 & 0.00 & 0.00 & 0.00 & 0.00 & 0.00 & 0.00  \\ \hline
			$ \mathcal{Y}^8 $ & 0.00 & 0.00 & 0.00 & 0.01 & 0.00 & -0.03 & 0.00 & 0.03 & 0.00 & -0.01  \\ \hline
		\end{tabular}

\caption{$N(1700)\tfrac{3}{2}^-$: quark+diquark orbital angular momentum breakdown of the canonical normalisation constant, drawn in Fig.\,\ref{LFigures}B.  The subarrays are composed according to the legend in Fig.\,\ref{LWFlegend} and the sum of all entries is unity. \label{tab:L-decompz-n1700-sc-av-dq}}
		\centering
		\begin{tabular}{c|c|cccc|cccc|c}\hline
			~ & $\mathcal{Y}^1$ & $ \mathcal{Y}^2 $ & $\mathcal{X}^1$ & $ \mathcal{Y}^3 $ & $ \mathcal{Y}^4 $ & $ \mathcal{Y}^5 $ & $ \mathcal{X}^2 $ & $ \mathcal{Y}^6 $ & $ \mathcal{Y}^7 $ & $ \mathcal{Y}^8 $ \\ \hline
			$\mathcal{Y}^1$ & -0.04 & -0.06 & -0.04 & 0.08 & 0.00 & 0.02 & 0.00 & 0.01 & 0.00 & -0.03  \\ \hline
			$ \mathcal{Y}^2 $ & -0.06 & 0.87 & 0.00 & 0.00 & 0.00 & 0.05 & -0.01 & -0.02 & 0.00 & 0.01  \\
			$\mathcal{X}^1$ & -0.04 & 0.00 & 0.21 & 0.08 & 0.00 & 0.00 & -0.23 & 0.00 & 0.00 & 0.00  \\
			$ \mathcal{Y}^3 $ & 0.08 & 0.00 & 0.08 & 0.08 & 0.00 & -0.02 & 0.00 & -0.01 & 0.00 & 0.00  \\
			$ \mathcal{Y}^4 $ & 0.00 & 0.00 & 0.00 & 0.00 & 0.00 & 0.00 & 0.00 & 0.00 & 0.00 & 0.00  \\ \hline
			$ \mathcal{Y}^5 $ & 0.02 & 0.05 & 0.00 & -0.02 & 0.00 & -0.04 & 0.00 & -0.01 & 0.00 & 0.09  \\
			$ \mathcal{X}^2 $ & 0.00 & -0.01 & -0.23 & 0.00 & 0.00 & 0.00 & 0.17 & -0.01 & 0.00 & 0.00  \\
			$ \mathcal{Y}^6 $ & 0.01 & -0.02 & 0.00 & -0.01 & 0.00 & -0.01 & -0.01 & -0.08 & 0.00 & 0.11  \\
			$ \mathcal{Y}^7 $ & 0.00 & 0.00 & 0.00 & 0.00 & 0.00 & 0.00 & 0.00 & 0.00 & 0.00 & 0.00  \\ \hline
			$ \mathcal{Y}^8 $ & -0.03 & 0.01 & 0.00 & 0.00 & 0.00 & 0.09 & 0.00 & 0.11 & 0.00 & -0.19  \\ \hline
		\end{tabular}
\end{table}

\newpage


	\begin{table}[th]
\caption{
$N(1720)\tfrac{3}{2}^+$: quark+diquark orbital angular momentum breakdown of the canonical normalisation constant, drawn in Fig.\,\ref{LFigures}C.  The subarrays are composed according to the legend in Fig.\,\ref{LWFlegend} and the sum of all entries is unity.  \label{tab:L-decompz-n1720-sc-av-dq}}
		\centering
		\begin{tabular}{c|c|cccc|cccc|c}
			\hline
			~ & $\mathcal{Y}^1$ & $ \mathcal{Y}^2 $ & $\mathcal{X}^1$ & $ \mathcal{Y}^3 $ & $ \mathcal{Y}^4 $ & $ \mathcal{Y}^5 $ & $ \mathcal{X}^2 $ & $ \mathcal{Y}^6 $ & $ \mathcal{Y}^7 $ & $ \mathcal{Y}^8 $ \\ \hline
			$\mathcal{Y}^1$ & -0.11 & -0.01 & 0.01 & 0.00 & 0.00 & 0.00 & 0.00 & -0.01 & 0.00 & 0.02  \\ \hline
			$ \mathcal{Y}^2 $ & -0.01 & 0.00 & 0.00 & 0.00 & 0.00 & -0.02 & 0.00 & 0.04 & 0.00 & -0.01  \\
			$\mathcal{X}^1$ & 0.01 & 0.00 & 0.10 & 0.00 & 0.00 & 0.00 & 0.04 & 0.00 & 0.00 & 0.00  \\
			$ \mathcal{Y}^3 $ & 0.00 & 0.00 & 0.00 & -0.01 & 0.00 & 0.00 & 0.00 & 0.01 & 0.00 & 0.00  \\
			$ \mathcal{Y}^4 $ & 0.00 & 0.00 & 0.00 & 0.00 & 0.00 & 0.00 & 0.00 & 0.00 & 0.00 & 0.00  \\ \hline
			$ \mathcal{Y}^5 $ & 0.00 & -0.02 & 0.00 & 0.00 & 0.00 & -0.04 & 0.00 & -0.02 & 0.00 & 0.12  \\
			$ \mathcal{X}^2 $ & 0.00 & 0.00 & 0.04 & 0.00 & 0.00 & 0.00 & 0.37 & 0.06 & 0.00 & 0.00  \\
			$ \mathcal{Y}^6 $ & -0.01 & 0.04 & 0.00 & 0.01 & 0.00 & -0.02 & 0.06 & -0.27 & 0.00 & 0.60  \\
			$ \mathcal{Y}^7 $ & 0.00 & 0.00 & 0.00 & 0.00 & 0.00 & 0.00 & 0.00 & 0.00 & 0.00 & 0.00  \\ \hline
			$ \mathcal{Y}^8 $ & 0.02 & -0.01 & 0.00 & 0.00 & 0.00 & 0.12 & 0.00 & 0.60 & 0.00 & -0.66  \\ \hline
		\end{tabular}


\caption{
$N(1900)\tfrac{3}{2}^+$: quark+diquark orbital angular momentum breakdown of the canonical normalisation constant, drawn in Fig.\,\ref{LFigures}D.  The subarrays are composed according to the legend in Fig.\,\ref{LWFlegend} and the sum of all entries is unity. \label{tab:L-decompz-n1900-sc-av-dq}}
		\centering
		\begin{tabular}{c|c|cccc|cccc|c}
			\hline
			~ & $\mathcal{Y}^1$ & $ \mathcal{Y}^2 $ & $\mathcal{X}^1$ & $ \mathcal{Y}^3 $ & $ \mathcal{Y}^4 $ & $ \mathcal{Y}^5 $ & $ \mathcal{X}^2 $ & $ \mathcal{Y}^6 $ & $ \mathcal{Y}^7 $ & $ \mathcal{Y}^8 $ \\ \hline
			$\mathcal{Y}^1$ & -0.32 & 0.05 & -0.04 & 0.35 & 0.00 & -0.02 & 0.00 & -0.01 & 0.00 & 0.03   \\ \hline
			$\mathcal{Y}^2$ & 0.05 & 0.27 & 0.00 & 0.00 & 0.00 & -0.01 & 0.02 & -0.07 & 0.00 & 0.02  \\
			$\mathcal{X}^1$ & -0.04 & 0.00 & 3.95 & 0.34 & 0.00 & -0.02 & -3.62 & -0.01 & 0.00 & 0.00   \\
			$\mathcal{Y}^3$ & 0.35 & 0.00 & 0.34 & -0.19 & 0.00 & 0.29 & -0.04 & -0.09 & 0.00 & -0.08   \\
			$\mathcal{Y}^4$ & 0.00 & 0.00 & 0.00 & 0.00 & 0.00 & 0.00 & 0.00 & 0.00 & 0.00 & 0.00  \\ \hline
			$\mathcal{Y}^5$ & -0.02 & -0.01 & -0.02 & 0.29 & 0.00 & -0.18 & 0.00 & 0.01 & 0.00 & -0.10   \\
			$\mathcal{X}^2$ & 0.00 & 0.02 & -3.62 & -0.04 & 0.00 & 0.00 & 3.71 & 0.05 & 0.00 & 0.02   \\
			$\mathcal{Y}^6$ & -0.01 & -0.07 & -0.01 & -0.09 & 0.00 & 0.01 & 0.05 & 0.60 & 0.00 & -1.18  \\
			$\mathcal{Y}^7$ & 0.00 & 0.00 & 0.00 & 0.00 & 0.00 & 0.00 & 0.00 & 0.00 & 0.00 & 0.00  \\ \hline
			$\mathcal{Y}^8$ & 0.03 & 0.02 & 0.00 & -0.08 & 0.00 & -0.10 & 0.02 & -1.18 & 0.00 & 1.40  \\ \hline
		\end{tabular}
	\end{table}


\begin{thebibliography}{88}
\providecommand{\natexlab}[1]{#1}
\providecommand{\url}[1]{\texttt{#1}}
\providecommand{\urlprefix}{URL }
\expandafter\ifx\csname urlstyle\endcsname\relax
  \providecommand{\doi}[1]{doi:\discretionary{}{}{}#1}\else
  \providecommand{\doi}[1]{doi:\discretionary{}{}{}\begingroup
  \urlstyle{rm}\url{#1}\endgroup}\fi
\providecommand{\bibinfo}[2]{#2}

\bibitem[{Brodsky et~al.(2022)Brodsky, Deur, and Roberts}]{Brodsky:2022fqy}
\bibinfo{author}{S.~J. Brodsky}, \bibinfo{author}{A.~Deur},
  \bibinfo{author}{C.~D. Roberts}, \bibinfo{title}{{Artificial Dynamical
  Effects in Quantum Field Theory}}, \bibinfo{journal}{Nature Reviews Physics}
  (\bibinfo{year}{2022}) \bibinfo{pages}{May}.

\bibitem[{Edwards et~al.(2011)Edwards, Dudek, Richards, and
  Wallace}]{Edwards:2011jj}
\bibinfo{author}{R.~G. Edwards}, \bibinfo{author}{J.~J. Dudek},
  \bibinfo{author}{D.~G. Richards}, \bibinfo{author}{S.~J. Wallace},
  \bibinfo{title}{{Excited state baryon spectroscopy from lattice QCD}},
  \bibinfo{journal}{Phys. Rev. D} \bibinfo{volume}{84} (\bibinfo{year}{2011})
  \bibinfo{pages}{074508}.

\bibitem[{Eichmann et~al.(2016{\natexlab{a}})Eichmann, Sanchis-Alepuz,
  Williams, Alkofer, and Fischer}]{Eichmann:2016yit}
\bibinfo{author}{G.~Eichmann}, \bibinfo{author}{H.~Sanchis-Alepuz},
  \bibinfo{author}{R.~Williams}, \bibinfo{author}{R.~Alkofer},
  \bibinfo{author}{C.~S. Fischer}, \bibinfo{title}{{Baryons as relativistic
  three-quark bound states}}, \bibinfo{journal}{Prog. Part. Nucl. Phys.}
  \bibinfo{volume}{91} (\bibinfo{year}{2016}{\natexlab{a}})
  \bibinfo{pages}{1--100}.

\bibitem[{Qin and Roberts(2020)}]{Qin:2020rad}
\bibinfo{author}{S.-X. Qin}, \bibinfo{author}{C.~D. Roberts},
  \bibinfo{title}{{Impressions of the Continuum Bound State Problem in QCD}},
  \bibinfo{journal}{Chin. Phys. Lett.}
  \bibinfo{volume}{37}~(\bibinfo{number}{12}) (\bibinfo{year}{2020})
  \bibinfo{pages}{121201}.

\bibitem[{Roberts et~al.(1992)Roberts, Williams, and Krein}]{Krein:1990sf}
\bibinfo{author}{C.~D. Roberts}, \bibinfo{author}{A.~G. Williams},
  \bibinfo{author}{G.~Krein}, \bibinfo{title}{{On the implications of
  confinement}}, \bibinfo{journal}{Int. J. Mod. Phys. A} \bibinfo{volume}{7}
  (\bibinfo{year}{1992}) \bibinfo{pages}{5607--5624}.

\bibitem[{Roberts(2008)}]{Roberts:2007ji}
\bibinfo{author}{C.~D. Roberts}, \bibinfo{title}{{Hadron Properties and
  Dyson-Schwinger Equations}}, \bibinfo{journal}{Prog. Part. Nucl. Phys.}
  \bibinfo{volume}{61} (\bibinfo{year}{2008}) \bibinfo{pages}{50--65}.

\bibitem[{Horn and Roberts(2016)}]{Horn:2016rip}
\bibinfo{author}{T.~Horn}, \bibinfo{author}{C.~D. Roberts},
  \bibinfo{title}{{The pion: an enigma within the Standard Model}},
  \bibinfo{journal}{J. Phys. G.} \bibinfo{volume}{43} (\bibinfo{year}{2016})
  \bibinfo{pages}{073001}.

\bibitem[{Aznauryan et~al.(2013)}]{Aznauryan:2012ba}
\bibinfo{author}{I.~G. Aznauryan}, et~al., \bibinfo{title}{{Studies of Nucleon
  Resonance Structure in Exclusive Meson Electroproduction}},
  \bibinfo{journal}{Int. J. Mod. Phys. E} \bibinfo{volume}{22}
  (\bibinfo{year}{2013}) \bibinfo{pages}{1330015}.

\bibitem[{Briscoe et~al.(2015)Briscoe, D{\"o}ring, Haberzettl, Manley, Naruki,
  Strakovsky, and Swanson}]{Briscoe:2015qia}
\bibinfo{author}{W.~J. Briscoe}, \bibinfo{author}{M.~D{\"o}ring},
  \bibinfo{author}{H.~Haberzettl}, \bibinfo{author}{D.~M. Manley},
  \bibinfo{author}{M.~Naruki}, \bibinfo{author}{I.~I. Strakovsky},
  \bibinfo{author}{E.~S. Swanson}, \bibinfo{title}{{Physics opportunities with
  meson beams}}, \bibinfo{journal}{Eur. Phys. J. A}
  \bibinfo{volume}{51}~(\bibinfo{number}{10}) (\bibinfo{year}{2015})
  \bibinfo{pages}{129}.

\bibitem[{Brodsky et~al.(2020)}]{Brodsky:2020vco}
\bibinfo{author}{S.~J. Brodsky}, et~al., \bibinfo{title}{{Strong QCD from
  Hadron Structure Experiments}}, \bibinfo{journal}{Int. J. Mod. Phys. E}
  \bibinfo{volume}{29}~(\bibinfo{number}{08}) (\bibinfo{year}{2020})
  \bibinfo{pages}{2030006}.

\bibitem[{Barabanov et~al.(2021)}]{Barabanov:2020jvn}
\bibinfo{author}{M.~Y. Barabanov}, et~al., \bibinfo{title}{{Diquark
  Correlations in Hadron Physics: Origin, Impact and Evidence}},
  \bibinfo{journal}{Prog. Part. Nucl. Phys.} \bibinfo{volume}{116}
  (\bibinfo{year}{2021}) \bibinfo{pages}{103835}.

\bibitem[{Munczek(1995)}]{Munczek:1994zz}
\bibinfo{author}{H.~J. Munczek}, \bibinfo{title}{{Dynamical chiral symmetry
  breaking, Goldstone's theorem and the consistency of the Schwinger-Dyson and
  Bethe-Salpeter Equations}}, \bibinfo{journal}{Phys. Rev. D}
  \bibinfo{volume}{52} (\bibinfo{year}{1995}) \bibinfo{pages}{4736--4740}.

\bibitem[{Bender et~al.(1996)Bender, Roberts, and von Smekal}]{Bender:1996bb}
\bibinfo{author}{A.~Bender}, \bibinfo{author}{C.~D. Roberts},
  \bibinfo{author}{L.~von Smekal}, \bibinfo{title}{{Goldstone Theorem and
  Diquark Confinement Beyond Rainbow- Ladder Approximation}},
  \bibinfo{journal}{Phys. Lett. B} \bibinfo{volume}{380} (\bibinfo{year}{1996})
  \bibinfo{pages}{7--12}.

\bibitem[{Qin et~al.(2014)Qin, Roberts, and Schmidt}]{Qin:2014vya}
\bibinfo{author}{S.-X. Qin}, \bibinfo{author}{C.~D. Roberts},
  \bibinfo{author}{S.~M. Schmidt}, \bibinfo{title}{{Ward-Green-Takahashi
  identities and the axial-vector vertex}}, \bibinfo{journal}{Phys. Lett. B}
  \bibinfo{volume}{733} (\bibinfo{year}{2014}) \bibinfo{pages}{202--208}.

\bibitem[{Binosi et~al.(2016)Binosi, Chang, Qin, Papavassiliou, and
  Roberts}]{Binosi:2016rxz}
\bibinfo{author}{D.~Binosi}, \bibinfo{author}{L.~Chang}, \bibinfo{author}{S.-X.
  Qin}, \bibinfo{author}{J.~Papavassiliou}, \bibinfo{author}{C.~D. Roberts},
  \bibinfo{title}{{Symmetry preserving truncations of the gap and
  Bethe-Salpeter equations}}, \bibinfo{journal}{Phys. Rev. D}
  \bibinfo{volume}{93} (\bibinfo{year}{2016}) \bibinfo{pages}{096010}.

\bibitem[{Eichmann et~al.(2010)Eichmann, Alkofer, Krassnigg, and
  Nicmorus}]{Eichmann:2009qa}
\bibinfo{author}{G.~Eichmann}, \bibinfo{author}{R.~Alkofer},
  \bibinfo{author}{A.~Krassnigg}, \bibinfo{author}{D.~Nicmorus},
  \bibinfo{title}{{Nucleon mass from a covariant three-quark Faddeev
  equation}}, \bibinfo{journal}{Phys. Rev. Lett.} \bibinfo{volume}{104}
  (\bibinfo{year}{2010}) \bibinfo{pages}{201601}.

\bibitem[{Sanchis-Alepuz et~al.(2011)Sanchis-Alepuz, Eichmann, Villalba-Chavez,
  and Alkofer}]{Sanchis-Alepuz:2011egq}
\bibinfo{author}{H.~Sanchis-Alepuz}, \bibinfo{author}{G.~Eichmann},
  \bibinfo{author}{S.~Villalba-Chavez}, \bibinfo{author}{R.~Alkofer},
  \bibinfo{title}{{Delta and Omega masses in a three-quark covariant Faddeev
  approach}}, \bibinfo{journal}{Phys. Rev. D} \bibinfo{volume}{84}
  (\bibinfo{year}{2011}) \bibinfo{pages}{096003}.

\bibitem[{Sanchis-Alepuz and Fischer(2014)}]{Sanchis-Alepuz:2014sca}
\bibinfo{author}{H.~Sanchis-Alepuz}, \bibinfo{author}{C.~S. Fischer},
  \bibinfo{title}{{Octet and Decuplet masses: a covariant three-body Faddeev
  calculation}}, \bibinfo{journal}{Phys. Rev. D} \bibinfo{volume}{90}
  (\bibinfo{year}{2014}) \bibinfo{pages}{096001}.

\bibitem[{Eichmann et~al.(2016{\natexlab{b}})Eichmann, Fischer, and
  Sanchis-Alepuz}]{Eichmann:2016hgl}
\bibinfo{author}{G.~Eichmann}, \bibinfo{author}{C.~S. Fischer},
  \bibinfo{author}{H.~Sanchis-Alepuz}, \bibinfo{title}{{Light baryons and their
  excitations}}, \bibinfo{journal}{Phys. Rev. D} \bibinfo{volume}{94}
  (\bibinfo{year}{2016}{\natexlab{b}}) \bibinfo{pages}{094033}.

\bibitem[{Qin et~al.(2018)Qin, Roberts, and Schmidt}]{Qin:2018dqp}
\bibinfo{author}{S.-X. Qin}, \bibinfo{author}{C.~D. Roberts},
  \bibinfo{author}{S.~M. Schmidt}, \bibinfo{title}{{Poincar{\'e}-covariant
  analysis of heavy-quark baryons}}, \bibinfo{journal}{Phys. Rev. D}
  \bibinfo{volume}{97} (\bibinfo{year}{2018}) \bibinfo{pages}{114017}.

\bibitem[{Qin et~al.(2019)Qin, Roberts, and Schmidt}]{Qin:2019hgk}
\bibinfo{author}{S.-X. Qin}, \bibinfo{author}{C.~D. Roberts},
  \bibinfo{author}{S.~M. Schmidt}, \bibinfo{title}{{Spectrum of light- and
  heavy-baryons}}, \bibinfo{journal}{Few Body Syst.} \bibinfo{volume}{60}
  (\bibinfo{year}{2019}) \bibinfo{pages}{26}.

\bibitem[{Eichmann et~al.(2008)Eichmann, Alkofer, Cloet, Krassnigg, and
  Roberts}]{Eichmann:2008ae}
\bibinfo{author}{G.~Eichmann}, \bibinfo{author}{R.~Alkofer},
  \bibinfo{author}{I.~C. Cloet}, \bibinfo{author}{A.~Krassnigg},
  \bibinfo{author}{C.~D. Roberts}, \bibinfo{title}{{Perspective on
  rainbow-ladder truncation}}, \bibinfo{journal}{Phys. Rev. C}
  \bibinfo{volume}{77} (\bibinfo{year}{2008}) \bibinfo{pages}{042202(R)}.

\bibitem[{Eichmann et~al.(2009)Eichmann, Cloet, Alkofer, Krassnigg, and
  Roberts}]{Eichmann:2008ef}
\bibinfo{author}{G.~Eichmann}, \bibinfo{author}{I.~C. Cloet},
  \bibinfo{author}{R.~Alkofer}, \bibinfo{author}{A.~Krassnigg},
  \bibinfo{author}{C.~D. Roberts}, \bibinfo{title}{{Toward unifying the
  description of meson and baryon properties}}, \bibinfo{journal}{Phys. Rev. C}
  \bibinfo{volume}{79} (\bibinfo{year}{2009}) \bibinfo{pages}{012202(R)}.

\bibitem[{Roberts et~al.(2011)Roberts, Chang, Cloet, and
  Roberts}]{Roberts:2011cf}
\bibinfo{author}{H.~L.~L. Roberts}, \bibinfo{author}{L.~Chang},
  \bibinfo{author}{I.~C. Cloet}, \bibinfo{author}{C.~D. Roberts},
  \bibinfo{title}{{Masses of ground and excited-state hadrons}},
  \bibinfo{journal}{Few Body Syst.} \bibinfo{volume}{51} (\bibinfo{year}{2011})
  \bibinfo{pages}{1--25}.

\bibitem[{Julia-Diaz et~al.(2007)Julia-Diaz, Lee, Matsuyama, and
  Sato}]{JuliaDiaz:2007kz}
\bibinfo{author}{B.~Julia-Diaz}, \bibinfo{author}{T.~S.~H. Lee},
  \bibinfo{author}{A.~Matsuyama}, \bibinfo{author}{T.~Sato},
  \bibinfo{title}{{Dynamical coupled-channel model of pi N scattering in the $
  W \leq 2$-GeV nucleon resonance region}}, \bibinfo{journal}{Phys. Rev. C}
  \bibinfo{volume}{76} (\bibinfo{year}{2007}) \bibinfo{pages}{065201}.

\bibitem[{Suzuki et~al.(2010)Suzuki, Julia-Diaz, Kamano, Lee, Matsuyama, and
  Sato}]{Suzuki:2009nj}
\bibinfo{author}{N.~Suzuki}, \bibinfo{author}{B.~Julia-Diaz},
  \bibinfo{author}{H.~Kamano}, \bibinfo{author}{T.~S.~H. Lee},
  \bibinfo{author}{A.~Matsuyama}, \bibinfo{author}{T.~Sato},
  \bibinfo{title}{{Disentangling the Dynamical Origin of P-11 Nucleon
  Resonances}}, \bibinfo{journal}{Phys. Rev. Lett.} \bibinfo{volume}{104}
  (\bibinfo{year}{2010}) \bibinfo{pages}{042302}.

\bibitem[{R{\"o}nchen et~al.(2013)R{\"o}nchen, D{\"o}ring, Huang, Haberzettl,
  Haidenbauer, Hanhart, Krewald, Meissner, and Nakayama}]{Ronchen:2012eg}
\bibinfo{author}{D.~R{\"o}nchen}, \bibinfo{author}{M.~D{\"o}ring},
  \bibinfo{author}{F.~Huang}, \bibinfo{author}{H.~Haberzettl},
  \bibinfo{author}{J.~Haidenbauer}, \bibinfo{author}{C.~Hanhart},
  \bibinfo{author}{S.~Krewald}, \bibinfo{author}{U.~G. Meissner},
  \bibinfo{author}{K.~Nakayama}, \bibinfo{title}{{Coupled-channel dynamics in
  the reactions $\pi N \to \pi N$, $\eta N$, $K\Lambda$, $K\Sigma$}},
  \bibinfo{journal}{Eur. Phys. J. A} \bibinfo{volume}{49}
  (\bibinfo{year}{2013}) \bibinfo{pages}{44}.

\bibitem[{Kamano et~al.(2013)Kamano, Nakamura, Lee, and Sato}]{Kamano:2013iva}
\bibinfo{author}{H.~Kamano}, \bibinfo{author}{S.~X. Nakamura},
  \bibinfo{author}{T.~S.~H. Lee}, \bibinfo{author}{T.~Sato},
  \bibinfo{title}{{Nucleon resonances within a dynamical coupled-channels model
  of $\pi N$ and $\gamma N$ reactions}}, \bibinfo{journal}{Phys. Rev. C}
  \bibinfo{volume}{88} (\bibinfo{year}{2013}) \bibinfo{pages}{035209}.

\bibitem[{Garc\'\i{}a-Tecocoatzi et~al.(2017)Garc\'\i{}a-Tecocoatzi, Bijker,
  Ferretti, and Santopinto}]{Garcia-Tecocoatzi:2016rcj}
\bibinfo{author}{H.~Garc\'\i{}a-Tecocoatzi}, \bibinfo{author}{R.~Bijker},
  \bibinfo{author}{J.~Ferretti}, \bibinfo{author}{E.~Santopinto},
  \bibinfo{title}{{Self-energies of octet and decuplet baryons due to the
  coupling to the baryon-meson continuum}}, \bibinfo{journal}{Eur. Phys. J. A}
  \bibinfo{volume}{53}~(\bibinfo{number}{6}) (\bibinfo{year}{2017})
  \bibinfo{pages}{115}.

\bibitem[{Cahill et~al.(1989)Cahill, Roberts, and Praschifka}]{Cahill:1988dx}
\bibinfo{author}{R.~T. Cahill}, \bibinfo{author}{C.~D. Roberts},
  \bibinfo{author}{J.~Praschifka}, \bibinfo{title}{{Baryon structure and QCD}},
  \bibinfo{journal}{Austral. J. Phys.} \bibinfo{volume}{42}
  (\bibinfo{year}{1989}) \bibinfo{pages}{129--145}.

\bibitem[{Burden et~al.(1989)Burden, Cahill, and Praschifka}]{Burden:1988dt}
\bibinfo{author}{C.~J. Burden}, \bibinfo{author}{R.~T. Cahill},
  \bibinfo{author}{J.~Praschifka}, \bibinfo{title}{{Baryon Structure and {QCD}:
  Nucleon Calculations}}, \bibinfo{journal}{Austral. J. Phys.}
  \bibinfo{volume}{42} (\bibinfo{year}{1989}) \bibinfo{pages}{147--159}.

\bibitem[{Reinhardt(1990)}]{Reinhardt:1989rw}
\bibinfo{author}{H.~Reinhardt}, \bibinfo{title}{{Hadronization of Quark Flavor
  Dynamics}}, \bibinfo{journal}{Phys. Lett. B} \bibinfo{volume}{244}
  (\bibinfo{year}{1990}) \bibinfo{pages}{316--326}.

\bibitem[{Efimov et~al.(1990)Efimov, Ivanov, and Lyubovitskij}]{Efimov:1990uz}
\bibinfo{author}{G.~V. Efimov}, \bibinfo{author}{M.~A. Ivanov},
  \bibinfo{author}{V.~E. Lyubovitskij}, \bibinfo{title}{{Quark - diquark
  approximation of the three quark structure of baryons in the quark
  confinement model}}, \bibinfo{journal}{Z. Phys. C} \bibinfo{volume}{47}
  (\bibinfo{year}{1990}) \bibinfo{pages}{583--594}.

\bibitem[{Chen et~al.(2018)Chen, El-Bennich, Roberts, Schmidt, Segovia, and
  Wan}]{Chen:2017pse}
\bibinfo{author}{C.~Chen}, \bibinfo{author}{B.~El-Bennich},
  \bibinfo{author}{C.~D. Roberts}, \bibinfo{author}{S.~M. Schmidt},
  \bibinfo{author}{J.~Segovia}, \bibinfo{author}{S.~Wan},
  \bibinfo{title}{{Structure of the nucleon's low-lying excitations}},
  \bibinfo{journal}{Phys. Rev. D} \bibinfo{volume}{97} (\bibinfo{year}{2018})
  \bibinfo{pages}{034016}.

\bibitem[{Burkert and Roberts(2019)}]{Burkert:2019bhp}
\bibinfo{author}{V.~D. Burkert}, \bibinfo{author}{C.~D. Roberts},
  \bibinfo{title}{{Colloquium: Roper resonance: Toward a solution to the
  fifty-year puzzle}}, \bibinfo{journal}{Rev. Mod. Phys.} \bibinfo{volume}{91}
  (\bibinfo{year}{2019}) \bibinfo{pages}{011003}.

\bibitem[{Roberts(2018)}]{Roberts:2018hpf}
\bibinfo{author}{C.~D. Roberts}, \bibinfo{title}{{N* Structure and Strong
  QCD}}, \bibinfo{journal}{Few Body Syst.} \bibinfo{volume}{59}
  (\bibinfo{year}{2018}) \bibinfo{pages}{72}.

\bibitem[{Sun et~al.(2020)}]{Sun:2019aem}
\bibinfo{author}{M.~Sun}, et~al., \bibinfo{title}{{Roper State from Overlap
  Fermions}}, \bibinfo{journal}{Phys. Rev. D} \bibinfo{volume}{101}
  (\bibinfo{year}{2020}) \bibinfo{pages}{054511}.

\bibitem[{Roberts et~al.(2013)Roberts, Holt, and Schmidt}]{Roberts:2013mja}
\bibinfo{author}{C.~D. Roberts}, \bibinfo{author}{R.~J. Holt},
  \bibinfo{author}{S.~M. Schmidt}, \bibinfo{title}{{Nucleon spin structure at
  very high $x$}}, \bibinfo{journal}{Phys. Lett. B} \bibinfo{volume}{727}
  (\bibinfo{year}{2013}) \bibinfo{pages}{249--254}.

\bibitem[{Cui et~al.(2020)Cui, Chen, Binosi, de~Soto, Roberts,
  Rodr{\'{\i}}guez-Quintero, Schmidt, and Segovia}]{Cui:2020rmu}
\bibinfo{author}{Z.-F. Cui}, \bibinfo{author}{C.~Chen},
  \bibinfo{author}{D.~Binosi}, \bibinfo{author}{F.~de~Soto},
  \bibinfo{author}{C.~D. Roberts},
  \bibinfo{author}{J.~Rodr{\'{\i}}guez-Quintero}, \bibinfo{author}{S.~M.
  Schmidt}, \bibinfo{author}{J.~Segovia}, \bibinfo{title}{{Nucleon elastic form
  factors at accessible large spacelike momenta}}, \bibinfo{journal}{Phys. Rev.
  D} \bibinfo{volume}{102} (\bibinfo{year}{2020}) \bibinfo{pages}{014043}.

\bibitem[{Chen et~al.(2022)Chen, Fischer, Roberts, and Segovia}]{Chen:2021guo}
\bibinfo{author}{C.~Chen}, \bibinfo{author}{C.~S. Fischer},
  \bibinfo{author}{C.~D. Roberts}, \bibinfo{author}{J.~Segovia},
  \bibinfo{title}{{Nucleon axial-vector and pseudoscalar form factors and PCAC
  relations}}, \bibinfo{journal}{Phys. Rev. D}
  \bibinfo{volume}{105}~(\bibinfo{number}{9}) (\bibinfo{year}{2022})
  \bibinfo{pages}{094022}.

\bibitem[{Cui et~al.(2022)Cui, Gao, Binosi, Chang, Roberts, and
  Schmidt}]{Cui:2021gzg}
\bibinfo{author}{Z.-F. Cui}, \bibinfo{author}{F.~Gao},
  \bibinfo{author}{D.~Binosi}, \bibinfo{author}{L.~Chang},
  \bibinfo{author}{C.~D. Roberts}, \bibinfo{author}{S.~M. Schmidt},
  \bibinfo{title}{{Valence quark ratio in the proton}}, \bibinfo{journal}{Chin.
  Phys. Lett. \emph{Express}} \bibinfo{volume}{39}~(\bibinfo{number}{04})
  (\bibinfo{year}{2022}) \bibinfo{pages}{041401}.

\bibitem[{Chang et~al.(2022)Chang, Gao, and Roberts}]{Chang:2022jri}
\bibinfo{author}{L.~Chang}, \bibinfo{author}{F.~Gao}, \bibinfo{author}{C.~D.
  Roberts}, \bibinfo{title}{{Parton distributions of light quarks and
  antiquarks in the proton}}, \bibinfo{journal}{Phys. Lett. B}
  \bibinfo{volume}{829} (\bibinfo{year}{2022}) \bibinfo{pages}{137078}.

\bibitem[{Cheng et~al.(2022)Cheng, Serna, Yao, Chen, Cui, and
  Roberts}]{Cheng:2022jxe}
\bibinfo{author}{P.~Cheng}, \bibinfo{author}{F.~E. Serna},
  \bibinfo{author}{Z.-Q. Yao}, \bibinfo{author}{C.~Chen},
  \bibinfo{author}{Z.-F. Cui}, \bibinfo{author}{C.~D. Roberts},
  \bibinfo{title}{{Contact interaction analysis of octet baryon axialvector and
  pseudoscalar form factors -- arXiv:2207.13811 [hep-ph]$\!$}} .

\bibitem[{Raya et~al.(2021)Raya, Guti\'errez-Guerrero, Bashir, Chang, Cui, Lu,
  Roberts, and Segovia}]{Raya:2021pyr}
\bibinfo{author}{K.~Raya}, \bibinfo{author}{L.~X. Guti\'errez-Guerrero},
  \bibinfo{author}{A.~Bashir}, \bibinfo{author}{L.~Chang},
  \bibinfo{author}{Z.~F. Cui}, \bibinfo{author}{Y.~Lu}, \bibinfo{author}{C.~D.
  Roberts}, \bibinfo{author}{J.~Segovia}, \bibinfo{title}{{Dynamical diquarks
  in the \mbox{$\gamma^{(\ast)} p\to N(1535)\tfrac{1}{2}^-$} transition}},
  \bibinfo{journal}{Eur. Phys. J. A} \bibinfo{volume}{57}~(\bibinfo{number}{9})
  (\bibinfo{year}{2021}) \bibinfo{pages}{266}.

\bibitem[{Crede and Roberts(2013)}]{Crede:2013kia}
\bibinfo{author}{V.~Crede}, \bibinfo{author}{W.~Roberts},
  \bibinfo{title}{{Progress towards understanding baryon resonances}},
  \bibinfo{journal}{Rept. Prog. Phys.} \bibinfo{volume}{76}
  (\bibinfo{year}{2013}) \bibinfo{pages}{076301}.

\bibitem[{Liu et~al.(2022)Liu, Chen, Lu, Roberts, and Segovia}]{Liu:2022ndb}
\bibinfo{author}{L.~Liu}, \bibinfo{author}{C.~Chen}, \bibinfo{author}{Y.~Lu},
  \bibinfo{author}{C.~D. Roberts}, \bibinfo{author}{J.~Segovia},
  \bibinfo{title}{{Composition of low-lying $J=\tfrac{3}{2}^\pm$
  \ensuremath{\Delta}-baryons}}, \bibinfo{journal}{Phys. Rev. D}
  \bibinfo{volume}{105}~(\bibinfo{number}{11}) (\bibinfo{year}{2022})
  \bibinfo{pages}{114047}.

\bibitem[{Workman et~al.(2022)}]{Workman:2022ynf}
\bibinfo{author}{R.~L. Workman}, et~al., \bibinfo{title}{{Review of Particle
  Physics}}, \bibinfo{journal}{PTEP} \bibinfo{volume}{2022}
  (\bibinfo{year}{2022}) \bibinfo{pages}{083C01}.

\bibitem[{Roberts(2020)}]{Roberts:2020hiw}
\bibinfo{author}{C.~D. Roberts}, \bibinfo{title}{{Empirical Consequences of
  Emergent Mass}}, \bibinfo{journal}{Symmetry} \bibinfo{volume}{12}
  (\bibinfo{year}{2020}) \bibinfo{pages}{1468}.

\bibitem[{Roberts(2021)}]{Roberts:2021xnz}
\bibinfo{author}{C.~D. Roberts}, \bibinfo{title}{{On Mass and Matter}},
  \bibinfo{journal}{AAPPS Bulletin} \bibinfo{volume}{31} (\bibinfo{year}{2021})
  \bibinfo{pages}{6}.

\bibitem[{Roberts et~al.(2021)Roberts, Richards, Horn, and
  Chang}]{Roberts:2021nhw}
\bibinfo{author}{C.~D. Roberts}, \bibinfo{author}{D.~G. Richards},
  \bibinfo{author}{T.~Horn}, \bibinfo{author}{L.~Chang},
  \bibinfo{title}{{Insights into the emergence of mass from studies of pion and
  kaon structure}}, \bibinfo{journal}{Prog. Part. Nucl. Phys.}
  \bibinfo{volume}{120} (\bibinfo{year}{2021}) \bibinfo{pages}{103883}.

\bibitem[{Binosi(2022)}]{Binosi:2022djx}
\bibinfo{author}{D.~Binosi}, \bibinfo{title}{{Emergent Hadron Mass in Strong
  Dynamics}}, \bibinfo{journal}{Few Body Syst.}
  \bibinfo{volume}{63}~(\bibinfo{number}{2}) (\bibinfo{year}{2022})
  \bibinfo{pages}{42}.

\bibitem[{Papavassiliou(2022)}]{Papavassiliou:2022wrb}
\bibinfo{author}{J.~Papavassiliou}, \bibinfo{title}{{Emergence of mass in the
  gauge sector of QCD -- arXiv:2207.04977 [hep-ph]}}, \bibinfo{journal}{Chin.
  Phys. C}  (\bibinfo{year}{2022}) \bibinfo{pages}{\emph{in
  press}.$\;$}\urlprefix\url{http://iopscience.iop.org/article/10.1088/1674-1137/ac84ca}.

\bibitem[{Mokeev and Carman(2022{\natexlab{a}})}]{Mokeev:2022xfo}
\bibinfo{author}{V.~I. Mokeev}, \bibinfo{author}{D.~S. Carman},
  \bibinfo{title}{{Photo- and Electrocouplings of Nucleon Resonances}},
  \bibinfo{journal}{Few Body Syst.} \bibinfo{volume}{63}~(\bibinfo{number}{3})
  (\bibinfo{year}{2022}{\natexlab{a}}) \bibinfo{pages}{59}.

\bibitem[{Denisov et~al.(2018)}]{Denisov:2018unj}
\bibinfo{author}{O.~Denisov}, et~al., \bibinfo{title}{{Letter of Intent (Draft
  2.0): A New QCD facility at the M2 beam line of the CERN SPS}} .

\bibitem[{Aoki et~al.(2021)}]{Aoki:2021cqa}
\bibinfo{author}{K.~Aoki}, et~al., \bibinfo{title}{{Extension of the J-PARC
  Hadron Experimental Facility: Third White Paper -- arXiv:2110.04462
  [nucl-ex]$\!$}} .

\bibitem[{Cloet et~al.(2009)Cloet, Eichmann, El-Bennich, Kl{\"a}hn, and
  Roberts}]{Cloet:2008re}
\bibinfo{author}{I.~C. Cloet}, \bibinfo{author}{G.~Eichmann},
  \bibinfo{author}{B.~El-Bennich}, \bibinfo{author}{T.~Kl{\"a}hn},
  \bibinfo{author}{C.~D. Roberts}, \bibinfo{title}{{Survey of nucleon
  electromagnetic form factors}}, \bibinfo{journal}{Few Body Syst.}
  \bibinfo{volume}{46} (\bibinfo{year}{2009}) \bibinfo{pages}{1--36}.

\bibitem[{Yin et~al.(2021)Yin, Cui, Roberts, and Segovia}]{Yin:2021uom}
\bibinfo{author}{P.-L. Yin}, \bibinfo{author}{Z.-F. Cui},
  \bibinfo{author}{C.~D. Roberts}, \bibinfo{author}{J.~Segovia},
  \bibinfo{title}{{Masses of positive- and negative-parity hadron
  ground-states, including those with heavy quarks}}, \bibinfo{journal}{Eur.
  Phys. J. C} \bibinfo{volume}{81}~(\bibinfo{number}{4}) (\bibinfo{year}{2021})
  \bibinfo{pages}{327}.

\bibitem[{Segovia et~al.(2014)Segovia, Cloet, Roberts, and
  Schmidt}]{Segovia:2014aza}
\bibinfo{author}{J.~Segovia}, \bibinfo{author}{I.~C. Cloet},
  \bibinfo{author}{C.~D. Roberts}, \bibinfo{author}{S.~M. Schmidt},
  \bibinfo{title}{{Nucleon and $\Delta$ elastic and transition form factors}},
  \bibinfo{journal}{Few Body Syst.} \bibinfo{volume}{55} (\bibinfo{year}{2014})
  \bibinfo{pages}{1185--1222}.

\bibitem[{Arp(1998)}]{Arpack}
\bibinfo{note}{R.~B.~Lehoucq, D.~C.~Sorensen and C.~Yang, {\em ARPACK Users'
  Guide: Solution of Large-Scale Eigenvalue Problems with Implicitly Restarted
  Arnoldi Methods\/} (Society for Industrial \& Applied Mathematics)},
  \bibinfo{year}{1998}.

\bibitem[{Qiu(2021)}]{SPECTRA}
\bibinfo{author}{Y.~Qiu}, \bibinfo{note}{{\em Sparse Eigenvalue Computation
  Toolkit as a Redesigned ARPACK (SPECTRA)\/}
  (https://spectralib.org/index.html)}, \bibinfo{year}{2021}.

\bibitem[{Hecht et~al.(2002)Hecht, Oettel, Roberts, Schmidt, Tandy, and
  Thomas}]{Hecht:2002ej}
\bibinfo{author}{M.~B. Hecht}, \bibinfo{author}{M.~Oettel},
  \bibinfo{author}{C.~D. Roberts}, \bibinfo{author}{S.~M. Schmidt},
  \bibinfo{author}{P.~C. Tandy}, \bibinfo{author}{A.~W. Thomas},
  \bibinfo{title}{{Nucleon mass and pion loops}}, \bibinfo{journal}{Phys. Rev.
  C} \bibinfo{volume}{65} (\bibinfo{year}{2002}) \bibinfo{pages}{055204}.

\bibitem[{Sanchis-Alepuz et~al.(2014)Sanchis-Alepuz, Fischer, and
  Kubrak}]{Sanchis-Alepuz:2014wea}
\bibinfo{author}{H.~Sanchis-Alepuz}, \bibinfo{author}{C.~S. Fischer},
  \bibinfo{author}{S.~Kubrak}, \bibinfo{title}{{Pion cloud effects on baryon
  masses}}, \bibinfo{journal}{Phys. Lett. B} \bibinfo{volume}{733}
  (\bibinfo{year}{2014}) \bibinfo{pages}{151--157}.

\bibitem[{Carman et~al.(2020)Carman, Joo, and Mokeev}]{Carman:2020qmb}
\bibinfo{author}{D.~Carman}, \bibinfo{author}{K.~Joo},
  \bibinfo{author}{V.~Mokeev}, \bibinfo{title}{{Strong QCD Insights from
  Excited Nucleon Structure Studies with CLAS and CLAS12}},
  \bibinfo{journal}{Few Body Syst.} \bibinfo{volume}{61} (\bibinfo{year}{2020})
  \bibinfo{pages}{29}.

\bibitem[{Mokeev and Carman(2022{\natexlab{b}})}]{Mokeev:2021dab}
\bibinfo{author}{V.~I. Mokeev}, \bibinfo{author}{D.~S. Carman},
  \bibinfo{title}{{New baryon states in exclusive meson
  photo-/electroproduction with CLAS}}, \bibinfo{journal}{Rev. Mex. Fis.
  Suppl.} \bibinfo{volume}{3}~(\bibinfo{number}{3})
  (\bibinfo{year}{2022}{\natexlab{b}}) \bibinfo{pages}{0308024}.

\bibitem[{Aznauryan et~al.(2009)}]{CLAS:2009ces}
\bibinfo{author}{I.~G. Aznauryan}, et~al., \bibinfo{title}{{Electroexcitation
  of nucleon resonances from CLAS data on single pion electroproduction}},
  \bibinfo{journal}{Phys. Rev. C} \bibinfo{volume}{80} (\bibinfo{year}{2009})
  \bibinfo{pages}{055203}.

\bibitem[{Mokeev et~al.(2016)}]{Mokeev:2015lda}
\bibinfo{author}{V.~I. Mokeev}, et~al., \bibinfo{title}{{New Results from the
  Studies of the $N(1440)1/2^+$, $N(1520)3/2^-$, and $\Delta(1620)1/2^-$
  Resonances in Exclusive $ep \to e'p' \pi^+ \pi^-$ Electroproduction with the
  CLAS Detector}}, \bibinfo{journal}{Phys. Rev. C} \bibinfo{volume}{93}
  (\bibinfo{year}{2016}) \bibinfo{pages}{025206}.

\bibitem[{Mokeev(2020)}]{Mokeev:2020vab}
\bibinfo{author}{V.~I. Mokeev}, \bibinfo{title}{{Two Pion Photo- and
  Electroproduction with CLAS}}, \bibinfo{journal}{EPJ Web Conf.}
  \bibinfo{volume}{241} (\bibinfo{year}{2020}) \bibinfo{pages}{03003}.

\bibitem[{Lu et~al.(2019)Lu, Chen, Cui, Roberts, Schmidt, Segovia, and
  Zong}]{Lu:2019bjs}
\bibinfo{author}{Y.~Lu}, \bibinfo{author}{C.~Chen}, \bibinfo{author}{Z.-F.
  Cui}, \bibinfo{author}{C.~D. Roberts}, \bibinfo{author}{S.~M. Schmidt},
  \bibinfo{author}{J.~Segovia}, \bibinfo{author}{H.~S. Zong},
  \bibinfo{title}{{Transition form factors: $\gamma^\ast + p \to \Delta(1232)$,
  $\Delta(1600)$}}, \bibinfo{journal}{Phys. Rev. D} \bibinfo{volume}{100}
  (\bibinfo{year}{2019}) \bibinfo{pages}{034001}.

\bibitem[{Mokeev(2022)}]{VictorMokeevPresentation}
\bibinfo{author}{V.~I. Mokeev}, \bibinfo{title}{{Insight into EHM from results
  on electroexcitation of $\Delta(1600)3/2^+$ resonance}}, in:
  \bibinfo{booktitle}{{Perceiving the Emergence of Hadron Mass through AMBER @
  CERN - VII}},
  \bibinfo{pages}{\href{https://indico.cern.ch/event/1145356/contributions/4850020/}{Contribution
  8}}, \bibinfo{year}{2022}.

\bibitem[{Mokeev et~al.(2020)}]{Mokeev:2020hhu}
\bibinfo{author}{V.~I. Mokeev}, et~al., \bibinfo{title}{{Evidence for the
  $N'(1720)3/2^+$ Nucleon Resonance from Combined Studies of CLAS $\pi^+\pi^-p$
  Photo- and Electroproduction Data}}, \bibinfo{journal}{Phys. Lett. B}
  \bibinfo{volume}{805} (\bibinfo{year}{2020}) \bibinfo{pages}{135457}.

\bibitem[{Chen et~al.(2019)Chen, Lu, Binosi, Roberts, Rodr\'\i{}guez-Quintero,
  and Segovia}]{Chen:2018nsg}
\bibinfo{author}{C.~Chen}, \bibinfo{author}{Y.~Lu},
  \bibinfo{author}{D.~Binosi}, \bibinfo{author}{C.~D. Roberts},
  \bibinfo{author}{J.~Rodr\'\i{}guez-Quintero}, \bibinfo{author}{J.~Segovia},
  \bibinfo{title}{{Nucleon-to-Roper electromagnetic transition form factors at
  large $Q^2$}}, \bibinfo{journal}{Phys. Rev. D} \bibinfo{volume}{99}
  (\bibinfo{year}{2019}) \bibinfo{pages}{034013}.

\bibitem[{Chen et~al.(2021)Chen, Fischer, Roberts, and Segovia}]{Chen:2020wuq}
\bibinfo{author}{C.~Chen}, \bibinfo{author}{C.~S. Fischer},
  \bibinfo{author}{C.~D. Roberts}, \bibinfo{author}{J.~Segovia},
  \bibinfo{title}{{Form Factors of the Nucleon Axial Current}},
  \bibinfo{journal}{Phys. Lett. B} \bibinfo{volume}{815} (\bibinfo{year}{2021})
  \bibinfo{pages}{136150}.

\bibitem[{Chen and Roberts(2022)}]{ChenChen:2022qpy}
\bibinfo{author}{C.~Chen}, \bibinfo{author}{C.~D. Roberts},
  \bibinfo{title}{{Nucleon axial form factor at large momentum transfers --
  arXiv:2206.12518 [hep-ph]\mbox{\rule{-1ex}{0ex}}}} .

\bibitem[{Lu et~al.(2022)Lu, Chang, Raya, Roberts, and
  Rodr\'\i{}guez-Quintero}]{Lu:2022cjx}
\bibinfo{author}{Y.~Lu}, \bibinfo{author}{L.~Chang}, \bibinfo{author}{K.~Raya},
  \bibinfo{author}{C.~D. Roberts},
  \bibinfo{author}{J.~Rodr\'\i{}guez-Quintero}, \bibinfo{title}{{Proton and
  pion distribution functions in counterpoint}}, \bibinfo{journal}{Phys. Lett.
  B} \bibinfo{volume}{830} (\bibinfo{year}{2022}) \bibinfo{pages}{137130}.

\bibitem[{Lane(1974)}]{Lane:1974he}
\bibinfo{author}{K.~D. Lane}, \bibinfo{title}{{Asymptotic Freedom and Goldstone
  Realization of Chiral Symmetry}}, \bibinfo{journal}{Phys. Rev. D}
  \bibinfo{volume}{10} (\bibinfo{year}{1974}) \bibinfo{pages}{2605}.

\bibitem[{Politzer(1976)}]{Politzer:1976tv}
\bibinfo{author}{H.~D. Politzer}, \bibinfo{title}{{Effective Quark Masses in
  the Chiral Limit}}, \bibinfo{journal}{Nucl. Phys. B} \bibinfo{volume}{117}
  (\bibinfo{year}{1976}) \bibinfo{pages}{397}.

\bibitem[{Pagels and Stokar(1979)}]{Pagels:1979hd}
\bibinfo{author}{H.~Pagels}, \bibinfo{author}{S.~Stokar}, \bibinfo{title}{{The
  Pion Decay Constant, Electromagnetic Form-Factor and Quark Electromagnetic
  Selfenergy in QCD}}, \bibinfo{journal}{Phys. Rev. D} \bibinfo{volume}{20}
  (\bibinfo{year}{1979}) \bibinfo{pages}{2947}.

\bibitem[{Cahill and Roberts(1985)}]{Cahill:1985mh}
\bibinfo{author}{R.~T. Cahill}, \bibinfo{author}{C.~D. Roberts},
  \bibinfo{title}{{Soliton Bag Models of Hadrons from QCD}},
  \bibinfo{journal}{Phys. Rev. D} \bibinfo{volume}{32} (\bibinfo{year}{1985})
  \bibinfo{pages}{2419}.

\bibitem[{Bashir et~al.(2012)}]{Bashir:2012fs}
\bibinfo{author}{A.~Bashir}, et~al., \bibinfo{title}{{Collective perspective on
  advances in Dyson-Schwinger Equation QCD}}, \bibinfo{journal}{Commun. Theor.
  Phys.} \bibinfo{volume}{58} (\bibinfo{year}{2012}) \bibinfo{pages}{79--134}.

\bibitem[{Roberts and Schmidt(2000)}]{Roberts:2000aa}
\bibinfo{author}{C.~D. Roberts}, \bibinfo{author}{S.~M. Schmidt},
  \bibinfo{title}{{Dyson-Schwinger equations: Density, temperature and
  continuum strong QCD}}, \bibinfo{journal}{Prog. Part. Nucl. Phys.}
  \bibinfo{volume}{45} (\bibinfo{year}{2000}) \bibinfo{pages}{S1--S103}.

\bibitem[{Fischer(2019)}]{Fischer:2018sdj}
\bibinfo{author}{C.~S. Fischer}, \bibinfo{title}{{QCD at finite temperature and
  chemical potential from Dyson--Schwinger equations}}, \bibinfo{journal}{Prog.
  Part. Nucl. Phys.} \bibinfo{volume}{105} (\bibinfo{year}{2019})
  \bibinfo{pages}{1--60}.

\bibitem[{Ebert et~al.(2011)Ebert, Faustov, and Galkin}]{Ebert:2011kk}
\bibinfo{author}{D.~Ebert}, \bibinfo{author}{R.~N. Faustov},
  \bibinfo{author}{V.~O. Galkin}, \bibinfo{title}{{Spectroscopy and Regge
  trajectories of heavy baryons in the relativistic quark-diquark picture}},
  \bibinfo{journal}{Phys. Rev. D} \bibinfo{volume}{84} (\bibinfo{year}{2011})
  \bibinfo{pages}{014025}.

\bibitem[{Ferretti(2019)}]{Ferretti:2019zyh}
\bibinfo{author}{J.~Ferretti}, \bibinfo{title}{{Effective Degrees of Freedom in
  Baryon and Meson Spectroscopy}}, \bibinfo{journal}{Few Body Syst.}
  \bibinfo{volume}{60} (\bibinfo{year}{2019}) \bibinfo{pages}{17}.

\bibitem[{Ebert et~al.(2002)Ebert, Faustov, Galkin, and
  Martynenko}]{Ebert:2002ig}
\bibinfo{author}{D.~Ebert}, \bibinfo{author}{R.~N. Faustov},
  \bibinfo{author}{V.~O. Galkin}, \bibinfo{author}{A.~P. Martynenko},
  \bibinfo{title}{{Mass spectra of doubly heavy baryons in the relativistic
  quark model}}, \bibinfo{journal}{Phys. Rev. D} \bibinfo{volume}{66}
  (\bibinfo{year}{2002}) \bibinfo{pages}{014008}.

\bibitem[{Zhang and Huang(2008)}]{Zhang:2008rt}
\bibinfo{author}{J.-R. Zhang}, \bibinfo{author}{M.-Q. Huang},
  \bibinfo{title}{{Doubly heavy baryons in QCD sum rules}},
  \bibinfo{journal}{Phys. Rev. D} \bibinfo{volume}{78} (\bibinfo{year}{2008})
  \bibinfo{pages}{094007}.

\bibitem[{Chen et~al.(2017)Chen, Chen, Liu, Liu, and Zhu}]{Chen:2016spr}
\bibinfo{author}{H.-X. Chen}, \bibinfo{author}{W.~Chen},
  \bibinfo{author}{X.~Liu}, \bibinfo{author}{Y.-R. Liu}, \bibinfo{author}{S.-L.
  Zhu}, \bibinfo{title}{{A review of the open charm and open bottom systems}},
  \bibinfo{journal}{Rept. Prog. Phys.} \bibinfo{volume}{80}
  (\bibinfo{year}{2017}) \bibinfo{pages}{076201}.

\bibitem[{Li et~al.(2020)Li, Chang, Qin, and Wang}]{Li:2019ekr}
\bibinfo{author}{Q.~Li}, \bibinfo{author}{C.-H. Chang}, \bibinfo{author}{S.-X.
  Qin}, \bibinfo{author}{G.-L. Wang}, \bibinfo{title}{{Mass spectra and wave
  functions of the doubly heavy baryons with $J^P=1^+$ heavy diquark cores}},
  \bibinfo{journal}{Chin. Phys. C} \bibinfo{volume}{44}~(\bibinfo{number}{1})
  (\bibinfo{year}{2020}) \bibinfo{pages}{013102}.

\bibitem[{Yin et~al.(2019)Yin, Chen, Krein, Roberts, Segovia, and
  Xu}]{Yin:2019bxe}
\bibinfo{author}{P.-L. Yin}, \bibinfo{author}{C.~Chen},
  \bibinfo{author}{G.~Krein}, \bibinfo{author}{C.~D. Roberts},
  \bibinfo{author}{J.~Segovia}, \bibinfo{author}{S.-S. Xu},
  \bibinfo{title}{{Masses of ground-state mesons and baryons, including those
  with heavy quarks}}, \bibinfo{journal}{Phys. Rev. D}
  \bibinfo{volume}{100}~(\bibinfo{number}{3}) (\bibinfo{year}{2019})
  \bibinfo{pages}{034008}.

\end{thebibliography}

\end{document}